\documentclass[a4paper,11pt]{article}
\usepackage{jcappub} 
\usepackage{lineno}

\usepackage{comment}
\usepackage{orcidlink}
\usepackage{multirow}
\usepackage{natbib}

\title{\boldmath The Cosmic Web in the DESI Early Data Release: A Probabilistic Environment Catalog}

\author[a]{Diana C. Zapata-Zuluaga\,\orcidlink{0009-0009-9847-1129},}
\author[b]{Sofía Guevara-Montoya\,\orcidlink{0000-0002-2193-3809},}
\author[c]{Valeria Torres-Gomez\,\orcidlink{0009-0004-9975-7093},}
\author[c]{Juliana Hernandez\,\orcidlink{0009-0000-5424-7023}}
\author[c,d]{and Jaime E. Forero-Romero\,\orcidlink{0000-0002-2890-3725}}

\affiliation[a]{\textit{Instituto de Física---FCEN, Universidad de Antioquia, Calle 67 No. 53-108, CP 050010, Medellín, Colombia}}

\affiliation[b]{\textit{Departamento de Física, Universidad Nacional de Colombia, Cra. 45 No. 26-85, CP 111321, Bogotá, Colombia}}

\affiliation[c]{\textit{Departamento de Física, Universidad de los Andes, Cra. 1 No. 18A-10, Edificio Ip, CP 111711, Bogotá, Colombia}}

\affiliation[d]{\textit{Observatorio Astronómico, Universidad de los Andes, Cra. 1 No. 18A-10, Edificio H, CP 111711, Bogotá, Colombia}}

\emailAdd{dianac.zapata@udea.edu.co}
\emailAdd{soguevaram@unal.edu.co}
\emailAdd{v.torresg23@uniandes.edu.co}
\emailAdd{j.hernandezh2@uniandes.edu.co}
\emailAdd{je.forero@uniandes.edu.co}

\abstract{We present the first public cosmic-web environment catalog built on any DESI data release.
Using \textsc{ASTRA} (\textit{Algorithm for Stochastic Topological
RAnking}), we classify each object in the DESI Early Data Release (EDR) into
\textit{void}, \textit{sheet}, \textit{filament}, or \textit{knot} by combining
observed positions with matched random catalogs, without reconstructing a continuous density field. 
We apply this method to the four DESI extragalactic tracers --- Bright
Galaxy Survey (BGS), Luminous Red Galaxies (LRG), Emission Line Galaxies (ELG), and quasars (QSO) --- across the 20 EDR rosettes ($\sim\!175~\mathrm{deg}^2$ total), running $100$ realizations per tracer-zone pair to derive per-object membership probabilities and classification entropies. 
We calibrate the classification thresholds using BGS as an anchor to match the volume-filling fractions reported for GAMA, and recover a physically consistent web morphology across all tracers. 
For BGS, the galaxy-count fractions and the environmental dependence of the star formation rate, which were not used in the calibration, are consistent with GAMA, COSMOS, and SDSS-based references, validating the method against established benchmarks. 
A normalized mutual information analysis on BGS reveals a clear dependence of the statistical associations between galaxy color, stellar mass, and specific star formation rate across environments.
These results provide a new observational baseline for galaxy evolution studies with DESI. 
All data products and the open-source pipeline are publicly available.
}
\keywords{cosmic web, redshift surveys, galaxy surveys, galaxy evolution, galaxy clustering}

\begin{document}
\maketitle
\flushbottom

\section{Introduction}
\label{sec:intro}

At the largest observable scales, matter in the Universe is not 
distributed uniformly. Instead, it forms a complex network of 
structures known as the \emph{cosmic web}: a system of voids, 
sheets, filaments, and knots that extends across hundreds of 
megaparsecs \citep{Geller1989Mapping, 
10.1111/j.1365-2966.2009.14885.x, Hoffman_2012, 
10.1093/mnras/stz1106}. Mapping the cosmic web is important 
for understanding galaxy evolution and cosmology 
\citep{10.1093/mnras/sts162, Wang_2021_a, Lavaux_2012, 
10.1093/mnras/stz1106, 10.1093/mnras/sty3118, Tanimura_2020}.

Many methods have been developed to classify the cosmic web, 
each based on a different definition of environment. 
Tensor-based methods such as T-Web and V-Web assign 
environments using the tidal tensor or the velocity-shear 
tensor, and their results depend on the choice of smoothing 
scale and threshold \citep{Hahn_2007, Hoffman_2012}. 
Topological and multiscale methods, by contrast, try to 
recover the web structure without choosing a single scale. 
Examples are DisPerSE, which uses persistent homology on 
Delaunay tessellations, and NEXUS/NEXUS+, which applies 
morphological filtering to density, tidal, and velocity 
fields \citep{10.1111/j.1365-2966.2011.18394.x, Cautun_2012, 
schaap2000continuousfieldsdiscretesamples}. Dynamical 
classifiers such as ORIGAMI define morphology through 
shell-crossing along orthogonal axes \citep{Falck_2012}. 
A systematic comparison of many of these methods shows that, 
despite their differences, they tend to agree on the broad 
picture of the web when applied to the same dataset 
\citep{Libeskind_2018}.

A central motivation for classifying the cosmic web is to 
understand how galaxy properties depend on large-scale 
environment. Early steps in this direction came from the 
Sloan Digital Sky Survey (SDSS), which provided the first 
large-area maps of the local Universe and enabled cosmic-web 
studies at relatively low galaxy number densities over 
thousands of square degrees \citep{York_2000, Tempel_2014}. 
These studies established the basic picture of environmental 
dependence: galaxies in denser structures tend to be redder, 
more massive, and less actively star-forming than those in 
underdense regions \citep{Peng_2010, Parente_2024}. At 
smaller angular scales, surveys such as GAMA and COSMOS 
offer a complementary view: their high number density of 
 targets over limited areas allows 
higher-resolution studies of the web and its galaxy 
populations. GAMA has characterized the galaxy luminosity 
function, stellar mass, star formation rate, and morphology 
across void, sheet, filament, and knot environments with 
well-controlled statistics at low redshift 
\citep{Eardley_2015}. 
COSMOS has extended this analysis to higher redshifts by exploiting its deep 
photometry, showing how galaxy populations change depending on whether they live in the field, in filaments, 
or in denser regions \citep{10.1093/mnras/stx3055, 
Darvish_2017}. Together, SDSS, GAMA, and COSMOS represent 
the main observational references against which any new 
cosmic-web classification should be tested.

The \textit{Dark Energy Spectroscopic Instrument} (DESI) 
opens a new opportunity to extend these studies to a much 
larger spectroscopic dataset. DESI is mapping the 
three-dimensional distribution of tens of millions of 
galaxies and quasars across $\sim\!14{,}000~\mathrm{deg}^2$, 
producing the largest spectroscopic maps of the Universe 
obtained so far \citep{desicollaboration2016desiexperimentisciencetargeting, 
2024}. Importantly, the DESI Survey Validation fields were 
chosen to overlap with well-characterized external surveys, 
including COSMOS and several GAMA fields 
\citep{https://doi.org/10.5281/zenodo.13308269}. This 
overlap makes it possible to compare DESI-based results 
directly with previous benchmark studies.

In this work, we apply the probabilistic classifier \textsc{ASTRA} (\textit{Algorithm 
for Stochastic Topological RAnking}) to the Large Scale Structure (LSS) catalogs of the DESI Early Data Release (EDR), covering 
20 rosette fields and the four extragalactic tracers: 
BGS, LRG, ELG, and QSO \citep{2024, 
ross2024constructionlargescalestructurecatalogs}. The 
\textsc{ASTRA} method classifies each galaxy into a void, 
sheet, filament, or knot environment by combining the 
observed galaxy positions with matched random catalogs, 
without reconstructing a continuous density field. By 
running 100 independent realizations per tracer-zone pair, 
it produces per-object membership probabilities that 
explicitly quantify classification uncertainty.

This work makes three contributions to the study of the cosmic web.
The first is a new public resource for the community: a ready-to-use 
catalog of cosmic-web environments built on the DESI EDR clustering-ready 
catalogs. 

The second contribution is a validation of the \textsc{ASTRA} method at 
low redshift. For the BGS sample, the galaxy count fractions, stellar 
mass fractions, and star-formation-rate gradients across environments, none 
of which enter the calibration, agree well with results from GAMA, COSMOS, 
and SDSS \citep{Eardley_2015, Darvish_2017, Parente_2024}. This consistency shows 
that \textsc{ASTRA} recovers physically meaningful environments directly 
from spectroscopic data.

Third, we present new results for BGS galaxies. Using 
normalized mutual information (NMI), we quantify the statistical 
associations between galaxy color, stellar mass, and specific star 
formation rate as a function of environment, providing a new observational baseline for testing galaxy evolution models  with DESI.

The paper is organized as follows. We describe the DESI 
instrument and the EDR data used in this work in 
\autoref{sec:desi_description}. The \textsc{ASTRA} 
classification method is presented in 
\autoref{sec:methodology}. The results include: visual 
maps of the web classification 
(\autoref{sec:visual_inspection}); the effect of survey 
boundaries and fiber-assignment completeness 
(\autoref{sec:boundary_effects}); classification 
uncertainty from entropy 
(\autoref{sec:Classification_Uncertainty}); web-type 
fractions for all tracers 
(\autoref{sec:web_type_fraction}); stellar mass 
distributions as a function of environment 
(\autoref{sec:stellar_mass}); and star formation rate 
analysis including the NMI study for BGS 
(\autoref{sec:sfr}). We discuss the broader implications 
of our results in \autoref{sec:discussion}, and we 
summarize our main conclusions in the last section. The 
full description of the public data products released on 
Zenodo is provided in \autoref{app:data_release}.

\section{The Dark Energy Spectroscopic Instrument}
\label{sec:desi_description}

\textit{The Dark Energy Spectroscopic Instrument} (DESI) is a wide-field, multi-object spectrograph installed at the Mayall 4-meter telescope at Kitt Peak National Observatory. It was built to map the large-scale structure of the Universe and to constrain cosmic expansion and structure formation \citep{levi2013desiexperimentwhitepapersnowmass, desicollaboration2016desiexperimentisciencetargeting}. The following description summarizes only the aspects directly relevant to this work; we refer the reader to \citep{desicollaboration2016desiexperimentisciencetargeting} and \citep{2024} for a complete account of the instrument and data products.

DESI releases calibrated spectra, redshift measurements, spectral classifications, and value-added catalogs (VACs) through staged public data releases. 
The Early Data Release (EDR) contains end-to-end products from Survey Validation (the ``One-Percent Survey''), including a suite of VACs that together form a self-consistent dataset
for early scientific analyses and method development \citep{2024}. 
The most recent public release is DESI DR1 \citep{desicollaboration2025datarelease1dark}.

Observations are organized in two programs: \emph{dark time}, used when the sky is dark, and \emph{bright time}, which allows efficient observations under higher sky backgrounds \citep{Myers_2023}. In dark time, DESI targets luminous red galaxies (LRGs; nominal target ranges $0.4<z<1.1$), emission-line galaxies (ELGs; $0.6<z<1.6$), and quasars (QSOs; $0.8<z<3.5$). In bright time, the survey covers the magnitude-limited Bright Galaxy Survey (BGS; peaking around $0.1<z<0.4$) and the Milky Way Survey (MWS) \citep{Adame_2025, desicollaboration2016desiexperimentisciencetargeting, Myers_2023, osti_1713302}. Together, these tracers cover a broad redshift range and allow powerful measurements of the cosmic web and other large-scale structure statistics \citep{desicollaboration2016desiexperimentisciencetargeting}.

Targets are selected from optical photometry in the $g,r,z$ bands from the DESI Legacy Imaging Surveys, combined with \emph{WISE} infrared imaging \citep{Dey_2019}. 
The \texttt{desitarget} pipeline builds the targeting bitmasks and unique identifiers (e.g., \texttt{TARGETID}) used consistently across all data releases, providing a reproducible link between photometric selection and spectroscopic samples \citep{Myers_2023}.

For large-scale structure studies, DESI provides clustering-ready catalogs (LSS catalogs) designed for two-point statistics and related measurements.
These catalogs include survey masks and systematic weights that correct for spatially varying effects caused by imaging depth, observing conditions, and fiber-assignment incompleteness \citep{ross2024constructionlargescalestructurecatalogs}.
Matched random catalogs are constructed to sample the same survey geometry as the data, with redshifts and weights drawn from the data catalogs to ensure consistency in the radial distribution \citep{Adame_2025}.

\subsection{Survey Geometry and the Rosette Strategy}
\label{sec:rosette}

The One-Percent Survey (SV3) is the final phase of DESI Survey Validation and the source of the data used in this work \citep{2024}. It was conducted from 2021 April to June and covered 20 fields using a \emph{rosette} strategy, in which 10--13 overlapping tiles are observed with centers offset in a circle of radius $0.12^\circ$ around a common pointing. 
This dense overlap produces a high-completeness core of $\sim\!6.5~\mathrm{deg}^2$ per field \citep{2024}. The geometric footprint of each rosette is a circle of radius $1.75^\circ$, or $9.6~\mathrm{deg}^2$; after the LSS masks, the full sky area covered by the SV3 LSS catalogs is $\sim\!175~\mathrm{deg}^2$ across the 20 rosettes, with an overall completeness of 94\% for BGS, 95\% for LRG, 86\% for ELG, and 98\% for QSO. These overall values average over the whole footprint, including the outer rim of each rosette where fewer tiles overlap and the completeness drops; in the body of the rosettes the completeness is higher (e.g.\ 0.92 for ELG between $0.5^\circ$ and $1.5^\circ$ from the center, see \autoref{sec:boundary_effects}).

The 20 rosette centers (\autoref{tab:sv3_rosette_centers}) were chosen to overlap well-studied external survey fields with extensive multiwavelength imaging and spectroscopy \citep{2024}. The selected fields include COSMOS (zone~0), two GAMA fields, G12 and G15 (zones~1, 2, 8--10, 17), GOODS-North (zone~3), and the Coma cluster (zones~4, 16), among others. This overlap is directly relevant for our work: the GAMA fields serve as the primary benchmark for calibrating the ASTRA classification thresholds (see \autoref{sec:methodology}), and the COSMOS field enables a direct comparison of our star formation rate results with previous environment studies at low redshift (see \autoref{sec:sfr}).

\begin{table}[ht]
\centering
\begin{tabular}{cccc}

\hline
\hline
Rosette & RA [deg] & Dec [deg] & Field \\ 
\hline
\multirow{2}{*}{SV3 R0}  & \multirow{2}{*}{$150.10$} & \multirow{2}{*}{$2.182$} 
& \textbf{COSMOS}, DES deep, LSST deep,\\ 
&  & & CFHTLS-D2 HSC ultradeep, VVDS-F10 \\
SV3 R1  & 179.60 &  0.000 & \textbf{GAMA} G12, KiDS-N\\
SV3 R2  & 183.10 &  0.000 & \textbf{GAMA} G12, KiDS-N \\
SV3 R3  & 189.90 & 61.800 & GOODS-North \\
SV3 R4  & 194.75 & 28.200 & Coma cluster \\
SV3 R5  & 210.00 &  5.000 & VVDS-F14 \\
SV3 R6  & 215.50 & 52.500 & DEEP2, CFHTLS-D3/W3 \\
SV3 R7  & 217.80 & 34.400 & Bootes NDWFS/AGES \\
SV3 R8  & 216.30 & -0.600 & \textbf{GAMA} G15, HSC DR2, KiDS-N \\
SV3 R9  & 219.80 & -0.600 & \textbf{GAMA} G15, HSC DR2, KiDS-N \\
SV3 R10 & 218.05 &  2.430 & \textbf{GAMA} G15, HSC DR2, KiDS-N \\
SV3 R11 & 242.75 & 54.980 & ELAIS N1, HSC deep field \\
SV3 R12 & 241.05 & 43.450 & HSC DR2 \\
SV3 R13 & 245.88 & 43.450 & HSC DR2 \\
SV3 R14 & 252.50 & 34.500 & XDEEP2 \\
SV3 R15 & 269.73 & 66.020 & Ecliptic pole, Euclid deep field \\
SV3 R16 & 194.75 & 24.700 & Coma cluster outskirts \\
SV3 R17 & 212.80 & -0.600 & \textbf{GAMA} G15, HSC DR2, KiDS-N \\
SV3 R18 & 269.73 & 62.520 & Near ecliptic pole \\
SV3 R19 & 236.10 & 43.450 & HSC DR2 \\
\hline
\end{tabular}
\caption{\small One-Percent Survey rosette centers (SV3) with their field association. RA and Dec refer to the center of the rosette, following the EDR convention \citep{2024}.}
\label{tab:sv3_rosette_centers}
\end{table}

\subsection{Input Data: Value Added Catalogs from the DESI Early Data Release}
\label{EDR}

In this work, we use the clustering-ready Large Scale Structure (LSS) Value Added Catalog (VAC) \footnote{\url{https://data.desi.lbl.gov/public/edr/vac/edr/lss/v2.0/}} from the DESI EDR \citep{2024}, focusing on its four extragalactic tracers: the \emph{Bright Galaxy Survey} (BGS), \emph{Luminous Red Galaxies} (LRG), \emph{Emission Line Galaxies} (ELG), and \emph{Quasars} (QSO). 
These catalogs are cut to reliable spectroscopic redshifts within the intended clustering range for each tracer, and include survey masks and completeness weights that correct for variations in fiber assignment, imaging conditions, and redshift success rates \citep{2024, ross2024constructionlargescalestructurecatalogs}. 
Each data catalog is accompanied by matched random catalogs that sample the same survey footprint, with redshifts drawn from the data to reproduce the weighted $\mathrm{d}N/\mathrm{d}z$. The randoms do not reproduce the spatial pattern of fiber-assignment completeness of the galaxies; the consequences for the classification are quantified in \autoref{sec:boundary_effects}. 
These randoms are a key ingredient of the ASTRA method (see \autoref{sec:ASTRA}). Comoving coordinates are computed with the Planck 2018 cosmology throughout this work.

\begin{figure}[htbp]
\begin{center} 
\includegraphics[width=0.7\textwidth]{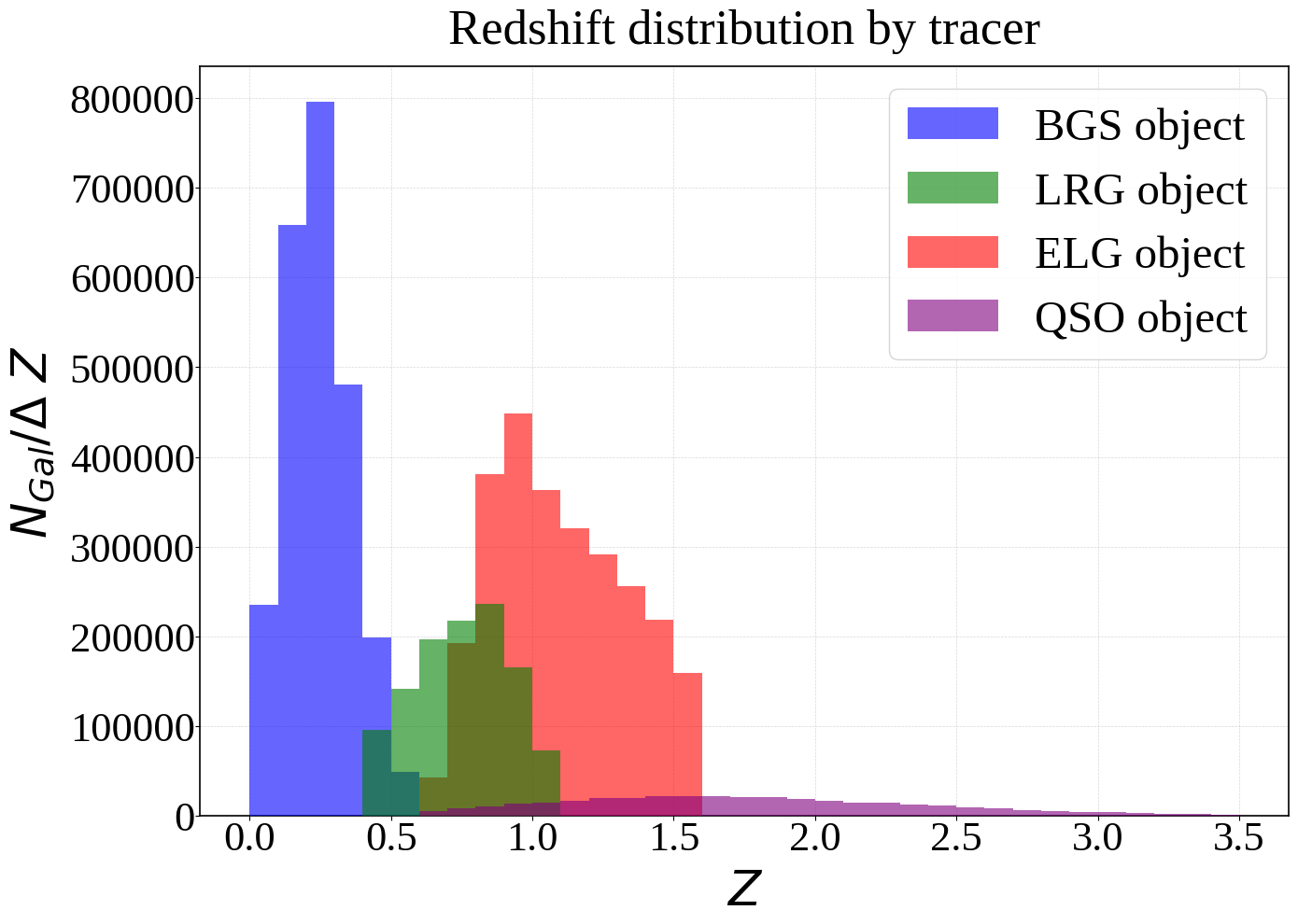}
\caption{\small{Redshift distribution of extragalactic objects in the DESI EDR, considering the full sample of the 20 zones (\emph{rosettes}) used in this work. The numbers of objects are 241,746, 112,649, 267,345, and 35,566 for BGS, LRG, ELG, and QSO, respectively.}}
\label{fig:data_z}
\end{center}
\end{figure}

The analysis is performed over the 20 EDR rosettes, which throughout this paper we denote as \textit{zone~0}--\textit{zone~19}. 
The ASTRA classification is carried out strictly by tracer and by zone; different populations and spatial regions are never mixed at that stage. Results that combine several zones or tracers, presented in \autoref{sec:results}, are stacked from these independent per-zone classifications. \autoref{fig:data_z} shows the redshift distribution of each tracer, illustrating their characteristic selection ranges and mutual overlap. 
The redshift intervals used are $0.01 < z < 0.6$ for BGS, $0.4 < z < 1.1$ for LRG, $0.6 < z < 1.6$ for ELG, and $0.6 < z < 3.5$ for QSO.
The number of objects in the EDR clustering catalogs is 241,746 for BGS, 112,649 for LRG, 267,345 for ELG, and 35,566 for QSO.

For the analysis of galaxy physical properties (Sections~\ref{sec:stellar_mass} and~\ref{sec:sfr}), we do not perform new SED fitting or emission-line modeling. Instead, we adopt the physical quantities provided by the DESI \texttt{stellar-mass-emline} VAC\footnote{\url{https://data.desi.lbl.gov/public/edr/vac/edr/stellar-mass-emline/v1.0/}}, which provides stellar masses and star formation rates estimated through spectral energy distribution (SED) fitting with CIGALE \citep{Boquien_2019}. The specific star formation rate is then computed as 
\texttt{SED\_SFR}/\texttt{SED\_MASS}. Therefore, throughout this work, stellar masses, SFRs, and derived sSFRs should be interpreted as catalog-level physical estimates rather than quantities newly inferred by our pipeline.

The physical parameters are inferred from optical-to-mid-infrared photometry and account for dust attenuation and possible AGN contribution \citep{Zou_2024}. We cross-match this VAC with the LSS catalogs using \texttt{TARGETID} and adopt the \texttt{SED\_MASS} and \texttt{SED\_SFR} columns. When a target has multiple entries due to repeated observations in different tiles, we retain the measurement with the smallest redshift uncertainty (\texttt{ZERR}) to ensure the most precise estimate is used. 

\section{Methodology} \label{sec:methodology}

\begin{figure}[htbp]
\begin{center} 
\includegraphics[width=0.7\textwidth]{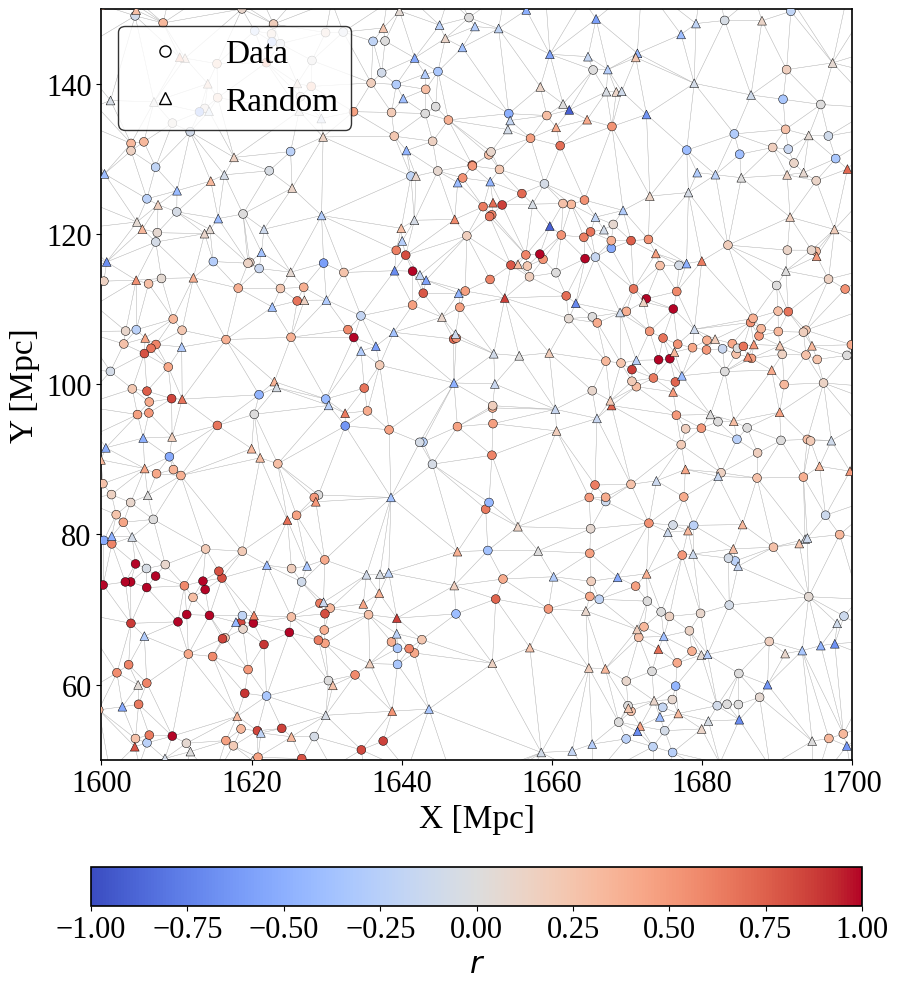}
\caption{\small{Example of the Delaunay tessellation built on the combined object and random catalog, $\mathcal{M}\equiv\mathcal{O}\cup\mathcal{R}$. Circles represent data points, triangles random points, and colors indicate the corresponding $r$ values.}}
\label{fig:ASTRA}
\end{center}
\end{figure}
\subsection{Algorithm for Stochastic Topological RAnking}
\label{sec:ASTRA}

\textsc{ASTRA} (\textit{Algorithm for Stochastic Topological RAnking}) probabilistically classifies the cosmic web using only the observed positions of galaxies and a matched random catalog that shares the same angular footprint and radial selection function \citep{foreroromero2025cosmicwebclassificationstochastic}. Unlike field-based classifiers, \textsc{ASTRA} operates directly on the discrete connectivity of a Delaunay graph, without reconstructing a continuous density field. The explicit use of random points with the same footprint and redshift distribution allows a joint characterization of overdense and underdense regions, including voids that are otherwise poorly traced by galaxies.
We reproduce here the key definitions from \citep{foreroromero2025cosmicwebclassificationstochastic} for self-containedness; readers familiar with that work may proceed directly to the recalibration in \autoref{tab:classification_adopted}.

\paragraph{The rank parameter $r$.}

The observed catalog \(\mathcal{O}\) and a matched random catalog \(\mathcal{R}\), with \(|\mathcal{R}| = |\mathcal{O}|\), are merged into \(\mathcal{M} \equiv \mathcal{O} \cup \mathcal{R}\), and a Delaunay tessellation is built on \(\mathcal{M}\). \autoref{fig:ASTRA} illustrates this tessellation for an example dataset. 

For each point \(p_i \in \mathcal{M}\), we count $N_{\mathcal{O}}(p_i)$ and $N_{\mathcal{R}}(p_i)$, the number of its Delaunay neighbors in \(\mathcal{O}\) and \(\mathcal{R}\) respectively, and define the rank parameter:

\begin{equation}
r(p_i) = \frac{N_{\mathcal{O}}(p_i) - N_{\mathcal{R}}(p_i)}{N_{\mathcal{O}}(p_i) + N_{\mathcal{R}}(p_i)} \in [-1, 1],
\label{eq:r_definition}
\end{equation}

where positive values indicate local overdensities and negative values indicate underdensities. Keeping $|\mathcal{R}| = |\mathcal{O}|$ establishes a consistent mean-density reference across the survey volume, so that the sign of $r$ directly separates overdense from underdense environments without requiring additional scaling factors \citep{foreroromero2025cosmicwebclassificationstochastic}. 
We denote by $\langle r \rangle$ the average of $r$ over the 100 random realizations we use in this paper.

\begin{figure}[htbp]
\begin{center}
\includegraphics[width=0.7\textwidth]{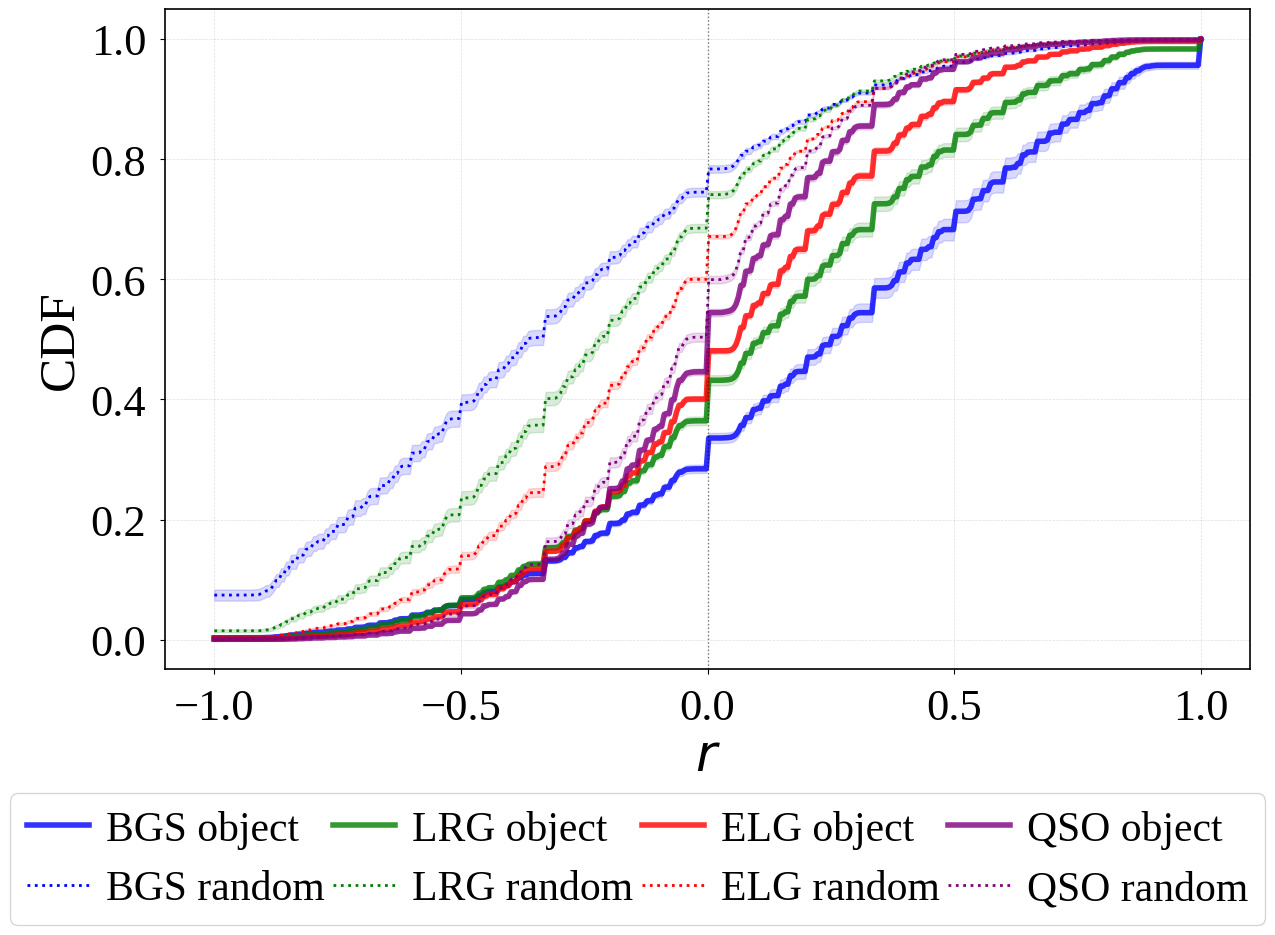}
\caption{\small Cumulative distribution function of $r$ for the real and random catalogs, separated by DESI tracer. Lines show the average over 100 realizations; shaded bands show the $1\sigma$ dispersion across the 20 zones.}
\label{fig:cdf}
\end{center}
\end{figure}

Each point is assigned to one of the four cosmic-web environments (\emph{void}, \emph{sheet}, \emph{filament}, \emph{knot}) by applying fixed thresholds on $r$. The original \textsc{ASTRA} thresholds are set conservatively at $|r| = 0.9$ to isolate only the most extreme density regimes \citep{foreroromero2025cosmicwebclassificationstochastic}, which yields void volume fractions well below those reported in the observational literature. We therefore recalibrate these thresholds to reproduce the volume-filling fractions (VFF) of GAMA \citep{Eardley_2015}, using the BGS tracer as anchor since it overlaps the GAMA redshift range. \autoref{fig:cdf} shows the cumulative distribution function (CDF) of $r$ for each DESI tracer, which guides this recalibration: for example, the original void threshold $r \leq -0.9$ gives a mean VFF of only $\sim 2.5\%$ across tracers, while the adopted threshold $r \leq -0.25$ gives $\sim 58\%$ for BGS, consistent with the GAMA value of $59\%$ at a smoothing scale of $4\,h^{-1}\,\mathrm{Mpc}$. The adopted thresholds are listed in \autoref{tab:classification_adopted}.

\begin{table}[htbp]
\centering
\begin{tabular}{cc}
\hline\hline
Condition & Classification \\
\hline
$-1 \leq r \leq -0.25$ & \emph{void} \\
$-0.25 < r \leq 0.25$ & \emph{sheet} \\
$0.25 < r \leq 0.65$ & \emph{filament} \\
$0.65 < r \leq 1$ & \emph{knot} \\
\hline
\end{tabular}
\caption{\small Adopted classification thresholds on $r$, recalibrated from the original \textsc{ASTRA} prescription \citep{foreroromero2025cosmicwebclassificationstochastic} to reproduce the GAMA volume-filling fractions \citep{Eardley_2015}.}
\label{tab:classification_adopted}
\end{table}

The distributions of $r$ shown in \autoref{fig:cdf} are smooth and do not display natural peaks associated with each web type. Therefore, the adopted thresholds should be understood as an operational calibration rather than as boundaries arising from distinct modes in the $r$ distribution. This situation is common to other cosmic-web classifiers. For example, in the T-Web, the eigenvalue threshold $\lambda_{\rm th}$ is also a free parameter that must be chosen a priori \citep{10.1111/j.1365-2966.2009.14885.x, Su_rez_P_rez_2021}. In this work, the thresholds are chosen to match the volume-filling fractions in the BGS tracer to those from the GAMA survey computed with the T-Web cosmic-web finder.

\begin{figure}[htbp] 
  \begin{center}
  \includegraphics[width=0.85\textwidth]{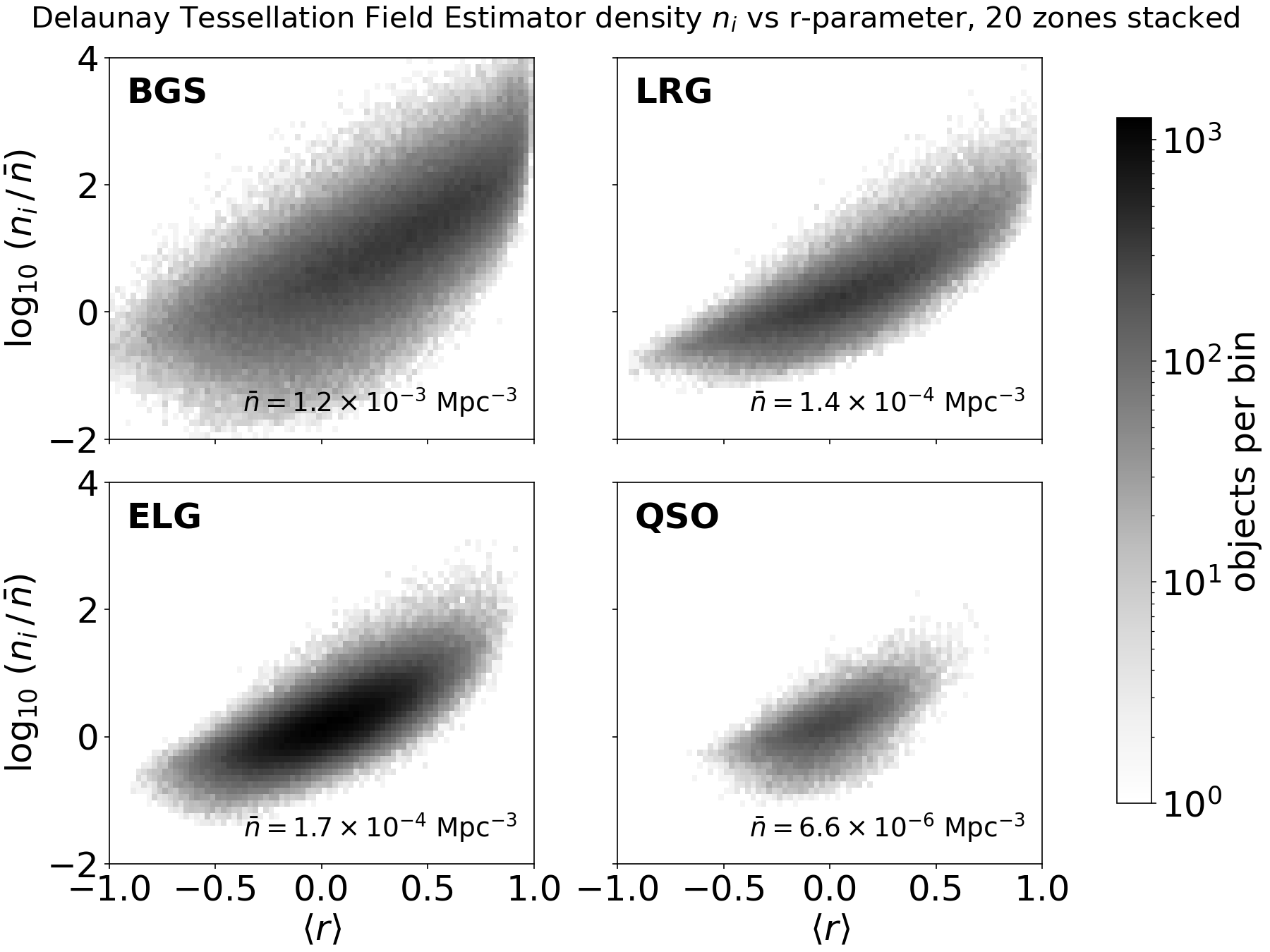}
  \caption{\small 2D histogram of the local number density $n_i$ against
  $\langle r \rangle $ for each tracer, with the 20 zones stacked. The density is the Delaunay
  Tessellation Field Estimator computed from the galaxies alone, without the
  random points. 
  Each panel is normalised by the mean density of the sample,
  $\bar{n} = N/V$, with $V$ the sum over the 20 zones of the comoving volume
  of the convex hull of the galaxies in each zone; the value of $\bar{n}$ is quoted in each panel. 
  The grey scale shows the number
  of galaxies per bin. The two quantities are correlated, but at fixed
  density a wide range of $\langle r\rangle$ values is allowed.}
  \label{fig:density_vs_r}
  \end{center}
  \end{figure}

Furthermore, to show that ASTRA is not simply a measurement of the local
  galaxy number density, we compute a local number density for every galaxy with the
  Delaunay Tessellation Field Estimator \citep{schaap2000continuousfieldsdiscretesamples}, $n_i = 4/W_i$,
  where $W_i$ is the total volume of the Delaunay tetrahedra that share
  galaxy $i$ as a vertex. 
  This tessellation is built from the galaxies alone: the
  random points that enter the definition of $r$ play no role in $n_i$,
  which is therefore a plain measurement of the galaxy density around each
  object. 
  \autoref{fig:density_vs_r} shows
  the 2D histogram of $n_i$, normalised by the mean density of each sample,
  against the mean ASTRA rank parameter $\langle r \rangle$, for all tracers
  with the 20 zones stacked. 
  The ridge of the distribution grows
  monotonically with $\langle r \rangle$, covering three orders of magnitude
  in density.
  However, the two quantities are loosely correlated: at fixed $\langle r \rangle$ the density
  spreads over one to two decades, and the spread is widest at the lowest
  values of $\langle r \rangle$. 
  Conversely, a single value of the local
  density is compatible with very different values of $\langle r \rangle$.

\paragraph{Probabilistic classification and uncertainty.}

To capture the sensitivity of the classification to the specific realization of the random catalog, we repeat the \textsc{ASTRA} workflow $N_{\mathrm{iter}} = 100$ times for each (tracer, zone) pair. In each iteration $k$, a different random subset $\mathcal{R}^{(k)}$ is drawn from the official EDR random catalogs, keeping $|\mathcal{R}^{(k)}| = |\mathcal{O}|$ and respecting the angular mask and radial selection. At the end of 100 runs, each real object has 100 class labels, from which we compute the membership probability for each environment $w$:

\begin{equation}
p_w = \frac{n_w}{100},
\end{equation}

where $n_w$ is the number of iterations in which the object was assigned to $w$. The classification uncertainty is quantified by the normalized Shannon entropy \citep{Shannon1949}:

\begin{equation}
H = -\frac{1}{\log_2 4} \sum_{w=1}^{4} p_w \log_2(p_w) \in [0,1],
\label{eq:entropia}
\end{equation}

where $H = 0$ means all iterations agreed on one class and $H = 1$ means all four classes were equally likely. A value $H \approx 0.5$ can arise in two ways: from an almost even split between two adjacent environments, or from a clearly dominant class that leaks a minor share of the probability to both of its neighboring environments (see \autoref{sec:Classification_Uncertainty}). In total, $4 \times 20 \times 100 = 8{,}000$ \textsc{ASTRA} runs are executed. The entropy results are analyzed in \autoref{sec:Classification_Uncertainty}.

\section{Results}
\label{sec:results}

\subsection{Cosmic-Web Classification Maps}
\label{sec:visual_inspection}

\begin{figure}[htbp]
\begin{center}
\includegraphics[width=0.9\textwidth]{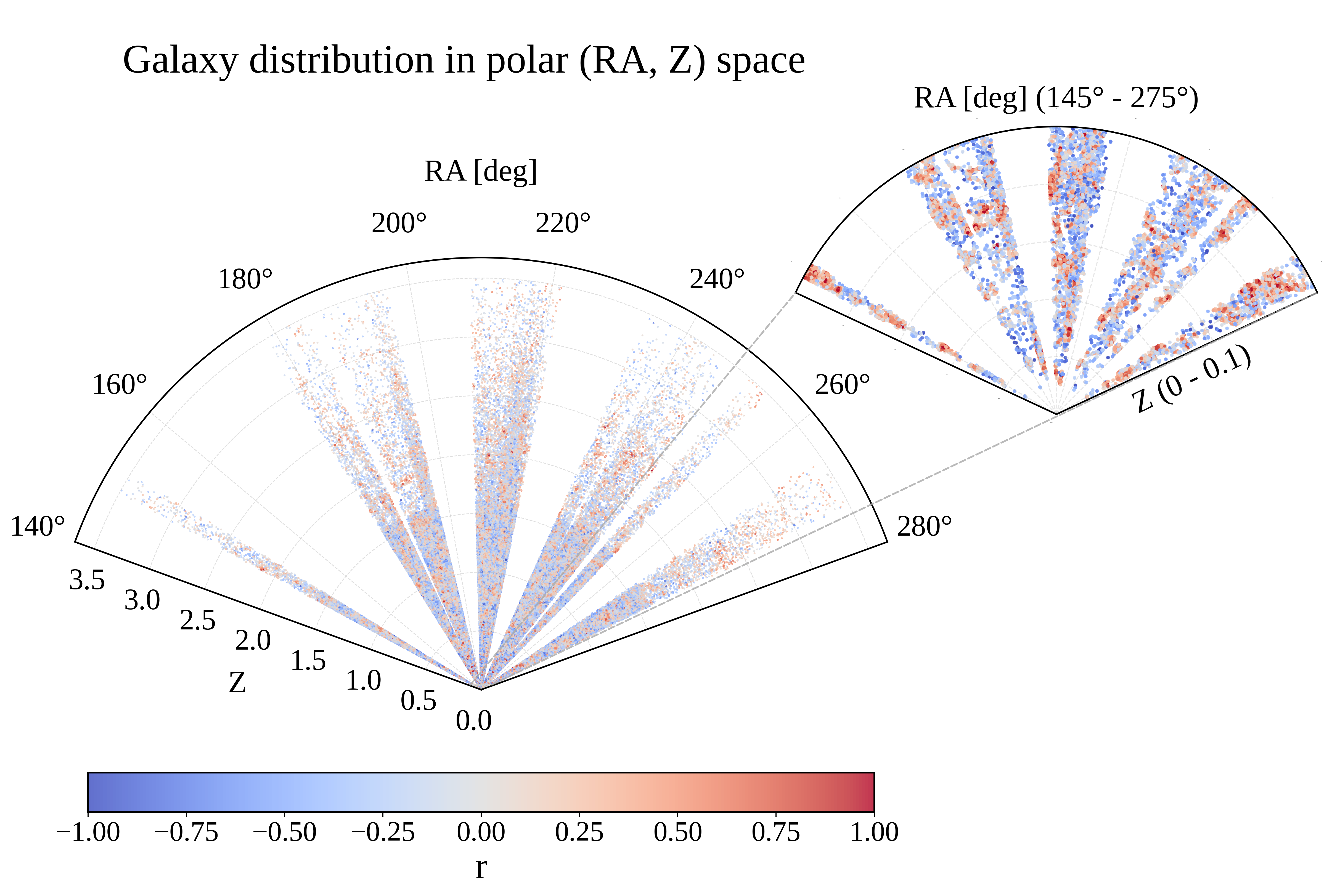}
\caption{\small Polar RA--$z$ distribution for the four extragalactic tracers across the 20 zones, colored by $r \in [-1,1]$. The inset ($\mathrm{RA} = 145^\circ$--$275^\circ$, $z = 0$--$0.1$) highlights overdense structures (filaments and knots) in reddish tones ($r > 0$) and underdense regions in bluish tones ($r < 0$); the sheet class straddles $r = 0$ (\autoref{tab:classification_adopted}).}
\label{fig:redshift_polar}
\end{center}
\end{figure}

\autoref{fig:redshift_polar} shows the full EDR sample across all 20 zones in a polar RA--$z$ projection, colored by $r \in [-1, 1]$. Reddish tones ($r > 0$) trace overdense structures, while bluish tones ($r < 0$) correspond to underdense regions. The inset ($\mathrm{RA} = 145^\circ$--$275^\circ$, $z = 0$--$0.1$) makes the filamentary pattern and its junctions clearly visible.

\begin{figure}[htbp]
\begin{center}
\includegraphics[width=1\textwidth]{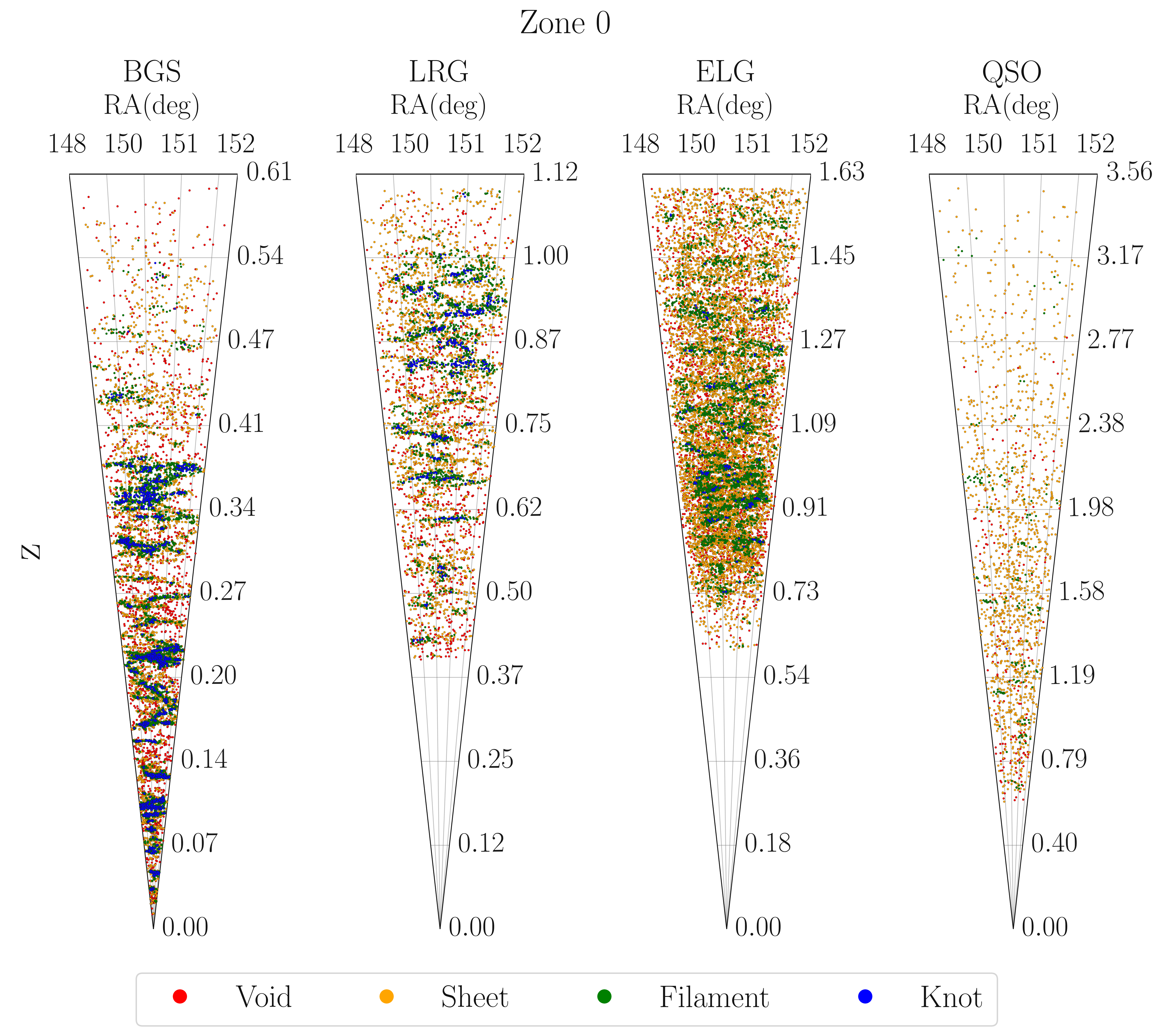}
\caption{\small Cone diagrams for zone~0, showing the most probable ASTRA classification for each real object across the four tracers.}
\label{fig:classification}
\end{center}
\end{figure}

\begin{figure}[htbp]
\begin{center}
\includegraphics[width=1\textwidth]{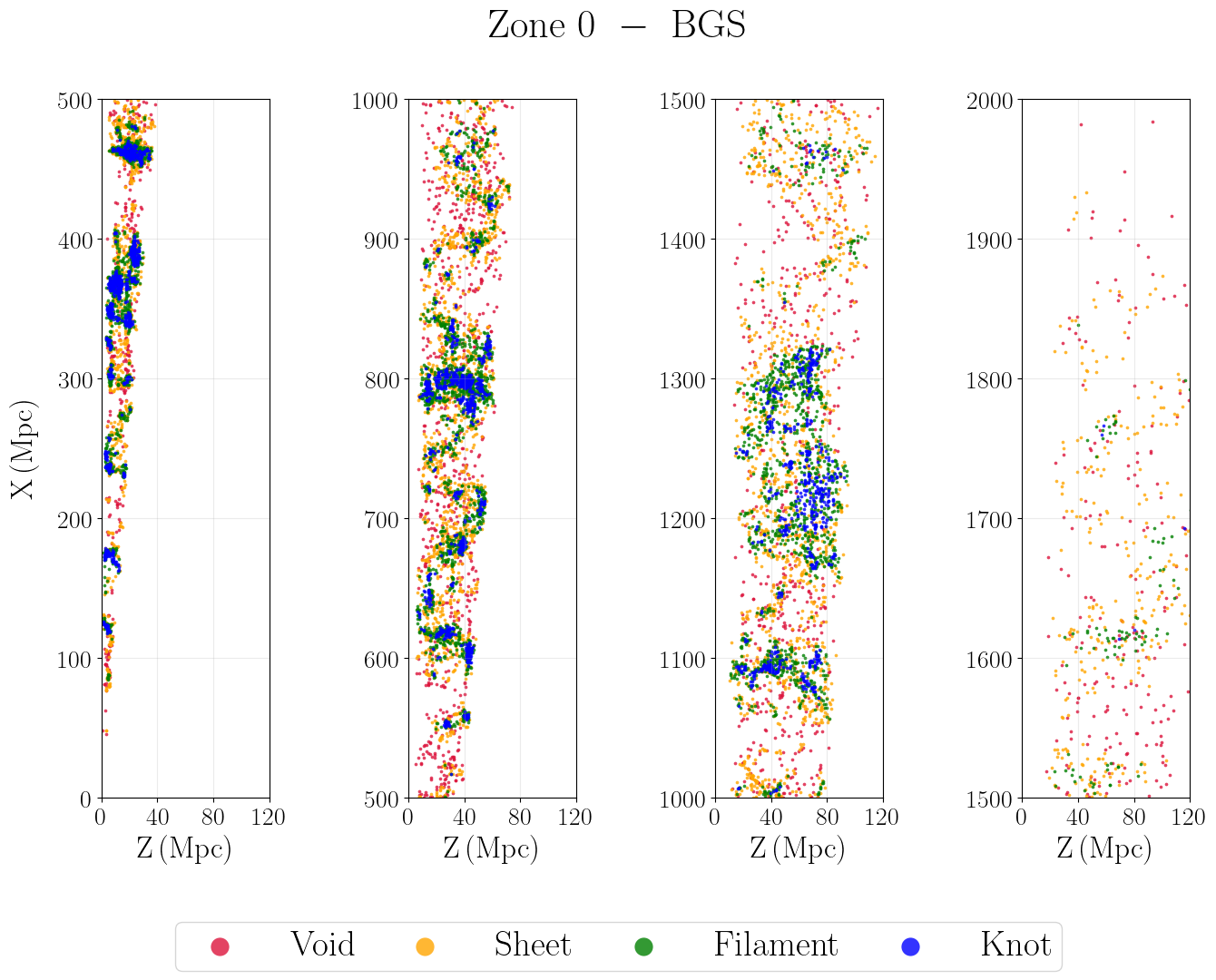}
\caption{\small Spatial distribution of BGS galaxies in Zone~0, classified by their most probable ASTRA cosmic-web type. The figure shows the \(X\)-\(Z\) projection in physical comoving coordinates with a strict 1:1 aspect ratio. For visualization, Zone~0 is divided into four consecutive \(X\)-ranges. The depth along the  \(Y\) direction is 120 Mpc approximately.}
\label{fig:wedge_zone_00_BGS}
\end{center}
\end{figure}

\autoref{fig:classification} shows cone diagrams for zone~0, with each point assigned its most probable web type over 100 ASTRA realizations. To complement this observational view, \autoref{fig:wedge_zone_00_BGS}
shows the BGS sample in physical comoving coordinates in the \(X\)-\(Z\) plane, using a 1:1 aspect ratio. Filaments trace anisotropic, elongated structures connecting overdense regions; sheets appear as more uniform, intermediate-density transition zones; knots emerge as sparse, point-like concentrations at the intersections of filaments; and voids are more frequent in less populated areas of the cone.

Going from BGS to QSO, the filamentary contrast decreases: BGS shows the sharpest structure, while QSO, being sparser and reaching higher $z$, shows a more diffuse pattern with fewer knots. This is consistent with the lower object density and higher sampling noise toward higher redshifts.

\subsection{Survey boundaries and fiber-assignment completeness}
\label{sec:boundary_effects}

\begin{figure}[htbp]
\begin{center} 
\includegraphics[width=1.\textwidth]{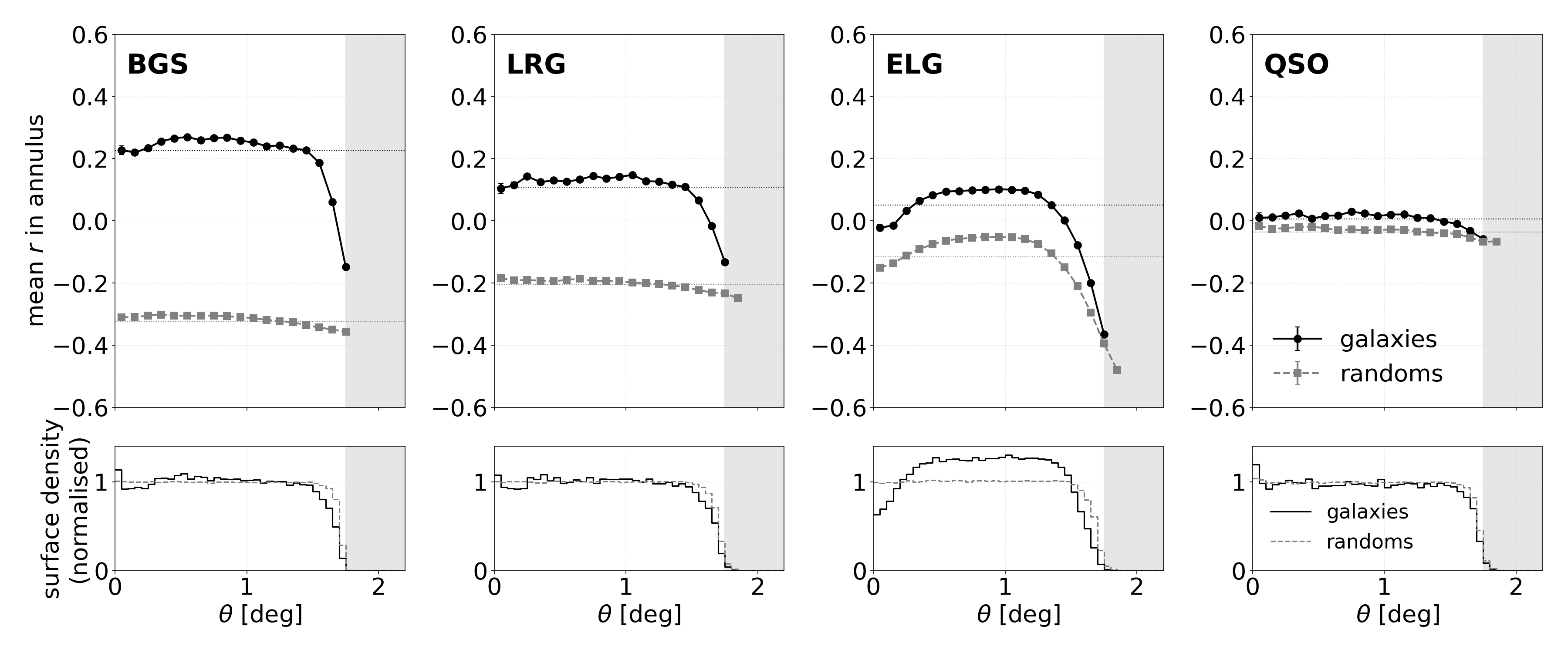}
\caption{\small{Top: mean $r$ in annuli of angular distance $\theta$ from the rosette center for the galaxies (black circles, $r$ averaged over the 100 realizations) and for the random points (grey squares, $r$ from individual random realizations, averaged over 100 realizations), stacked over the 20 rosettes. Error bars show the standard error of the mean and dotted lines the global means. Bottom: surface density of galaxies and randoms in the same annuli, normalized to the inner $0.5^\circ$. The grey band marks the region beyond the footprint edge at $\theta = 1.75^\circ$, where the random density falls below half its central value. The mean $r$ is flat for $\theta < 1.5^\circ$ and decreases only where the galaxy density drops relative to the randoms because of the lower fiber-assignment completeness of the rim and, for ELG, of
the rosette center.}}
\label{fig:rparam}
\end{center}
\end{figure}

The parameter $r$ is constructed from the Delaunay neighbors of each object and may therefore be affected by the edges of the survey footprint, where neighboring points are absent on one side. The use of random catalogs is intended to prevent large biases of this kind, since galaxies and random points share the same footprint and are similarly affected by the loss of neighbors near the boundary. To quantify any residual effect, we take advantage of the fact that all 20 EDR rosettes share the same circular geometry: a footprint of radius $1.75^\circ$ around the rosette center. Figure~\ref{fig:rparam} shows the mean $r$ of galaxies and random points in annuli of angular distance $\theta$ from the rosette center, stacked over the 20 rosettes, together with the corresponding surface-density profiles of both catalogs.

The DESI One-Percent Survey was specifically designed to obtain highly complete coverage in 20 rosette fields through strongly overlapping observations, with 10--13 overlapping tiles per rosette (\autoref{sec:rosette}) \citep{2024}. Thus, unlike the outer boundary of each rosette, internal survey holes are not a prominent geometrical feature of the footprint considered here. The relevant spatial variations within the rosettes are instead associated with changes in tile overlap and fiber-assignment completeness, as discussed below.

For all tracers, the mean $r$ remains approximately flat out to $\theta \simeq 1.5^\circ$, corresponding to the inner $73\%$ of the footprint area, indicating that the tessellation itself does not introduce a significant boundary bias over most of the rosette. In the outer rim, $1.5^\circ < \theta < 1.75^\circ$, the mean $r$ of the galaxies decreases, from $+0.25$ to $-0.15$ for BGS, $+0.13$ to $-0.13$ for LRG, $+0.08$ to $-0.37$ for ELG, and $+0.01$ to $-0.06$ for QSO in the outermost $0.1^\circ$, while for BGS, LRG, and QSO the mean $r$ of the randoms shows a much smaller decrease; for ELG, the randoms fall together with the galaxies in the rim. The ELG sample also exhibits a smaller decrease in $r$ within $0.3^\circ$ of the rosette center.

The bottom row of Figure~\ref{fig:rparam} identifies the origin of this behavior. The surface density of the random points remains approximately flat up to the edge of the footprint, whereas the galaxy surface density decreases in the outer rim and, for ELG, also near the rosette center. 
Both regions, center and outer rim, are covered by fewer overlapping tiles than the main body of the rosette. 
The mean number of DESI tiles covering a given position is 4--5 in the rim and 6--7 at the center, compared with 8--9.5 in between. 
This occurs because the tile centers are offset by only $0.12^\circ$, while the central region is always observed through the center of the focal plane. 
Fewer overlapping tiles result in lower fiber-assignment completeness, with the ELG sample being the most affected because it is the densest and lowest-priority dark-time target class \citep{2024}. 

In the EDR LSS catalogs, the fiber-assignment completeness (\texttt{COMP\_TILE}) sampled at the positions of the random points is $0.74$ within $0.3^\circ$ of the center and $0.51$ in the rim, compared with $0.92$ in the plateau at $0.5^\circ$--$1.5^\circ$. Relative to the plateau, this predicts an ELG number density of $80\%$ at the center and $55\%$ in the rim; the measured values are $78\%$ and $54\%$, so the completeness accounts for the deficit. The DESI LSS catalogs correct for this incompleteness by weighting each galaxy by the inverse of its completeness, whereas the random catalogs retain the full footprint density and only record the completeness associated with each position. Since ASTRA uses positions only, it inherits the spatial completeness pattern present in the galaxy catalog but not in the random catalog, causing $r$ to be biased toward lower values wherever the completeness decreases.

In this case the affected regions are well bounded by the radial coordinate. 
The outer rim, $\theta > 1.5^\circ$, contains $17.5\%$ of the BGS, $18.3\%$ of the LRG, $12.0\%$ of the ELG, and $21.9\%$ of the QSO objects, while the central region, $\theta < 0.3^\circ$, contains only about $3\%$ of the ELG sample. We provide the angular distance to the rosette center in the released catalog so that users can identify and exclude these regions when needed. The environmental trends presented in this work are dominated by the inner regions of the rosettes; therefore, throughout this work we restrict the analysis to $\theta \leq 1.5^\circ$ for BGS, LRG, and QSO, and to $0.3^\circ \leq \theta \leq 1.5^\circ$ for ELG. This restriction is applied to every table and figure in the paper, including those of \autoref{sec:methodology}; the only exception is \autoref{fig:rparam}, which by construction shows the full rosette. The released catalogs are not restricted.

A complete correction is straightforward as the random catalogs carry the completeness associated with each position. Subsampling the randoms with probability \texttt{COMP\_TILE} before running ASTRA ensures that both catalogs follow the same selection function. We will adopt this procedure for the analysis of the DESI DR1 catalogs. The radial test presented here relies on the common circular geometry of the EDR rosettes and cannot be directly applied to the irregular footprint of the main survey. For future data releases, the classification should instead be accompanied by a HEALPix map of the fiber-assignment completeness, constructed from the random catalogs, so that the dependence of $r$ and of the web-type fractions on completeness can be diagnosed directly on the sky.


\subsection{Classification Uncertainty}
\label{sec:Classification_Uncertainty}

\begin{figure}[htbp]
\begin{center}
\includegraphics[width=0.7\textwidth]{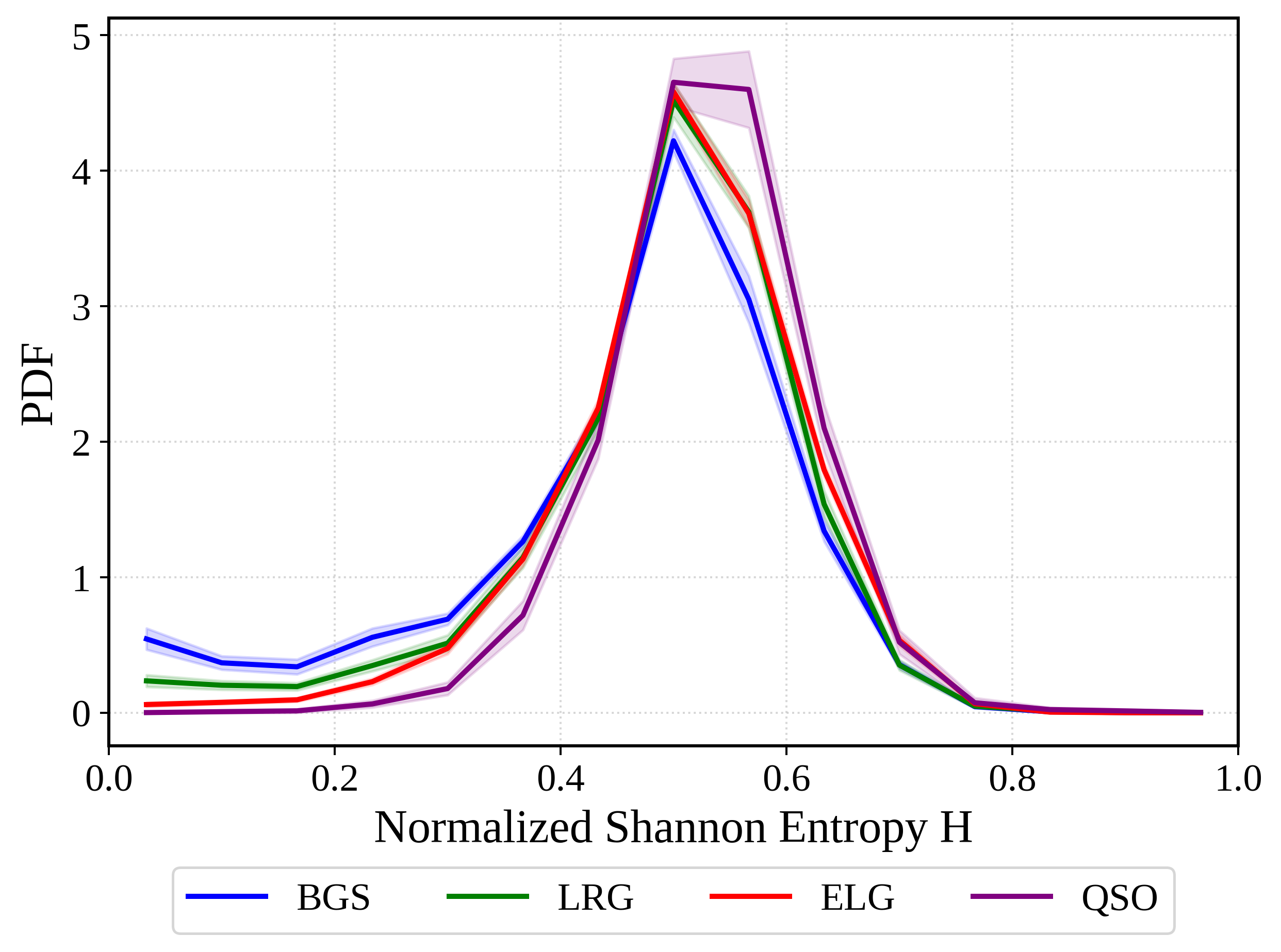}
\caption{\small PDF of the normalized entropy $H$ for each tracer. Curves show the average across the 20 zones; shaded bands show the $1\sigma$ inter-zone dispersion. All tracers peak near $H \sim 0.5$.}
\label{fig:entropy}
\end{center}
\end{figure}

\autoref{fig:entropy} shows the probability density function (PDF) of the normalized entropy $H$ (defined in \autoref{eq:entropia}) for each tracer, averaged over the 20 zones, with $1\sigma$ inter-zone bands.

All tracers peak near $H \sim 0.5$. As shown below, this peak is not produced by objects split evenly between two environments; it is dominated by sheets and filaments that keep a clearly dominant class while leaking a minor share of the probability to both of their neighboring environments. BGS shows the lowest peak entropy ($ H = 0.49$), consistent with having the highest comoving number density of the four tracers and therefore producing the most stable Delaunay triangulations. QSO shows the highest ($ H = 0.53$), reflecting its sparser sampling and greater sensitivity to changes in the random catalog.

\begin{figure}[htbp]
\begin{center}
\includegraphics[width=0.95\textwidth]{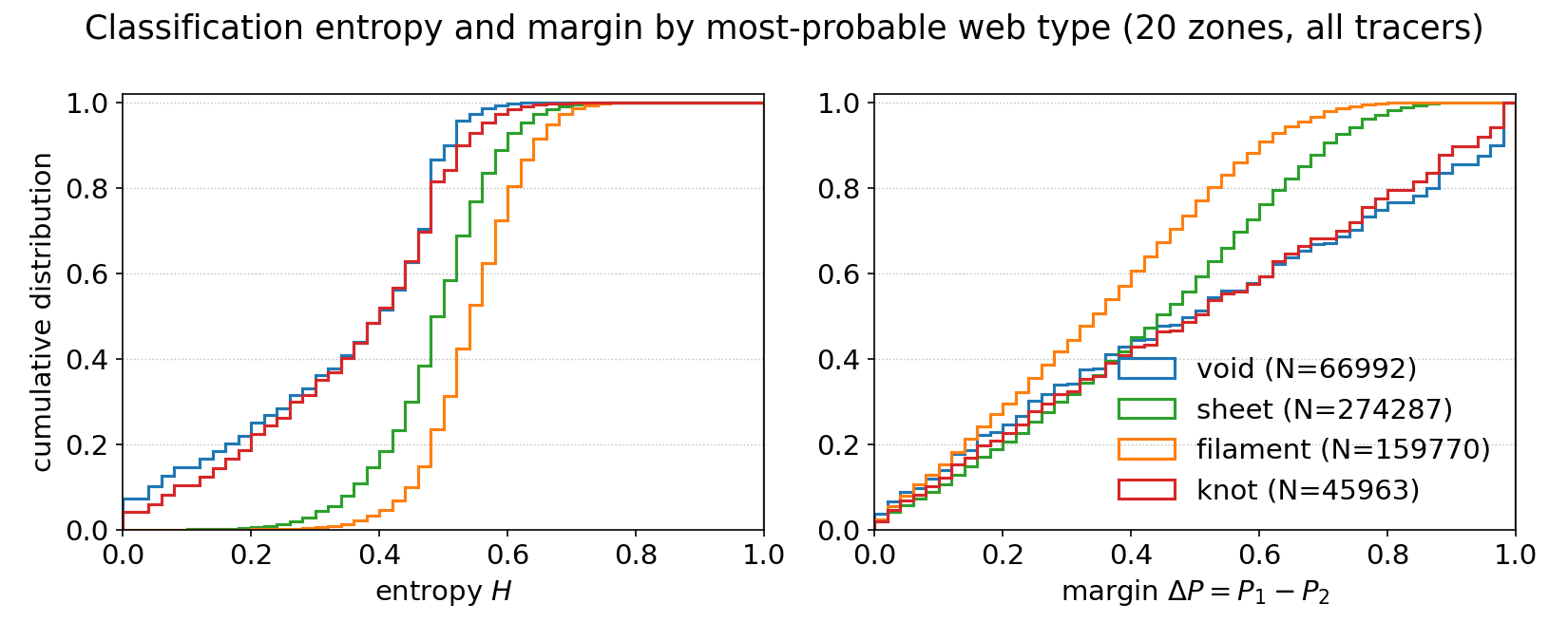}
\caption{\small Cumulative distributions of the classification entropy $H$ (left) and of the probability margin $\Delta P = P_1 - P_2$ (right), split by most-probable web type, for the 20 zones and the four tracers combined.}
\label{fig:entropy_margin}
\end{center}
\end{figure}

To understand how this uncertainty depends on the assigned environment, the entropy distribution is examined after splitting the sample by the most-probable web type. The left panel of \autoref{fig:entropy_margin} shows the cumulative distribution of \(H\), combining the 20 zones and the four tracers. The peak at \(H \approx 0.5\) is dominated by the intermediate environments, which make up the bulk of the sample: sheets and filaments together account for 434,057  of the 547,012  classified objects that remain after the angular cut of \autoref{sec:boundary_effects}, with median entropies of \(H=0.50\) and \(H=0.56\), respectively. Voids and knots show systematically lower entropies, both with median \(H \approx 0.41\), and extended tails towards \(H=0\): about 20\% of them have \(H<0.2\), while essentially no sheet or filament does.

The right panel of \autoref{fig:entropy_margin} shows the cumulative distribution of \(\Delta P = P_1 - P_2\), where \(P_1\) is the probability of the most-probable web type and \(P_2\) is the probability of the second most-probable web type for each galaxy. The margin gives a result consistent with the entropy: filaments are the least certain class, with median \(\Delta P=0.35\), followed by sheets, with median \(\Delta P=0.46\). Voids and knots have the largest margins, with median values of \(\Delta P=0.51\) and \(\Delta P=0.50\), respectively, and prominent high-confidence tails, with roughly 20\% of them at \(\Delta P>0.9\).

\begin{figure}[htbp]
\begin{center}
\includegraphics[width=0.7\textwidth]{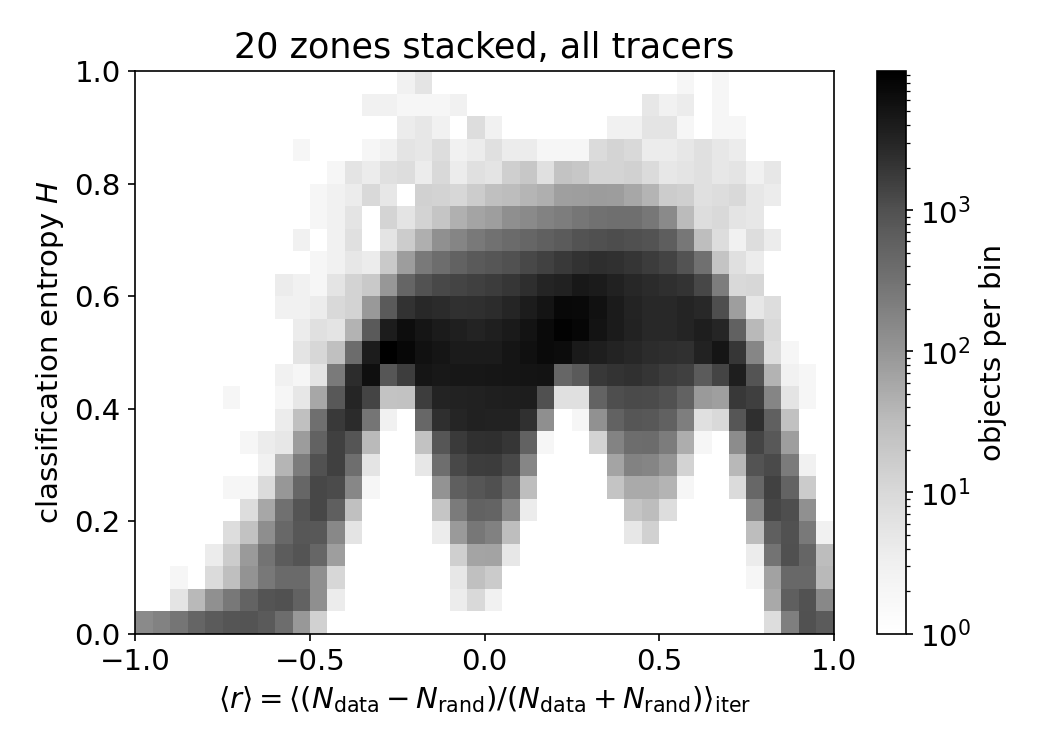}
\caption{\small 2D histogram of the classification entropy $H$ against $\langle r \rangle$ for the 20 zones and the four tracers combined. The entropy is lowest at the extremes of the $ r $ range and in the cores of the intermediate-type intervals, and highest near the classification boundaries.}
\label{fig:entropy_vs_r}
\end{center}
\end{figure}

The dependence of the entropy on the ASTRA parameter itself is shown in \autoref{fig:entropy_vs_r}. The entropy is lowest at the two extremes of the \( r\) range, where voids (\( r \to -1\)) and knots (\( r \to +1\)) are identified with high confidence. It also drops in the cores of the intermediate \( r\) intervals, where the sheet and filament assignments are stable. The highest entropies, \(H \approx 0.6\)--$0.8$, concentrate near the classification boundaries, where an object can scatter between neighboring types across the random iterations.

\begin{figure}[htbp]
\begin{center}
\includegraphics[width=0.7\textwidth]{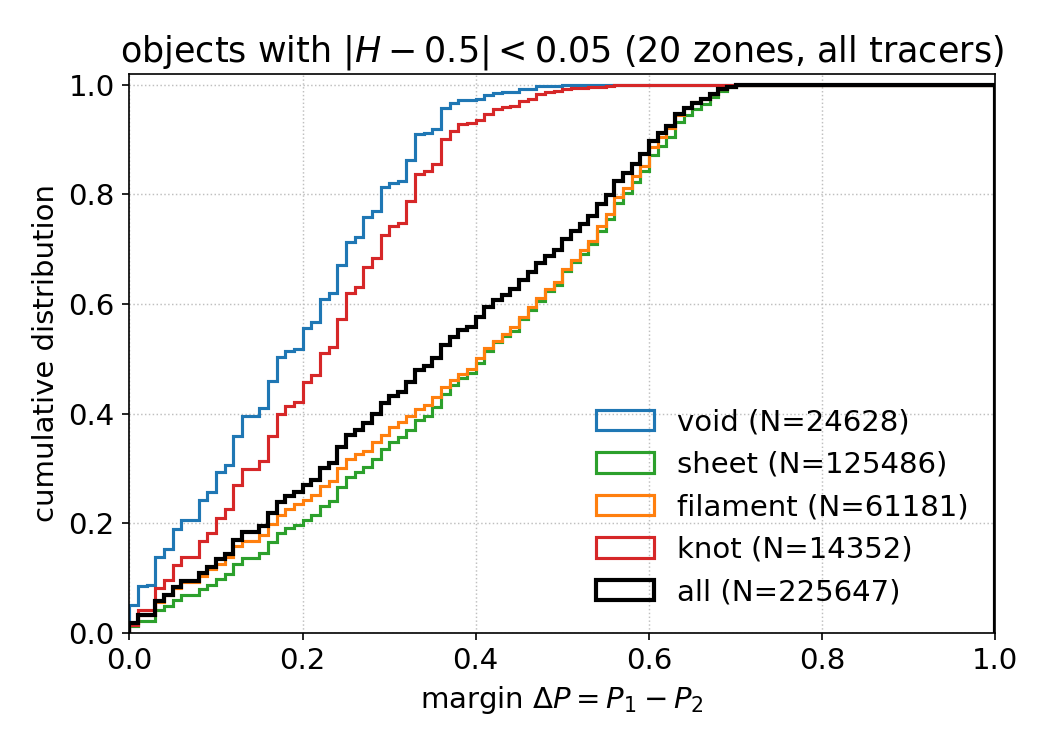}
\caption{\small Cumulative distribution of the margin $\Delta P = P_1 - P_2$ for the objects inside the entropy peak, $|H - 0.5| < 0.05$, split by most-probable web type (20 zones, all tracers combined). Even at fixed $H \approx 0.5$, sheets and filaments keep a sizeable margin, while the few voids and knots in this regime are genuinely ambiguous. No object exceeds $\Delta P \approx 0.7$, the limit imposed by the entropy selection itself.}
\label{fig:margin_H05}
\end{center}
\end{figure}

The objects inside the entropy peak itself are then examined, defining the peak region as \(|H-0.5|<0.05\). This selection contains 225,647 objects, about 41\% of the classified sample. \autoref{fig:margin_H05} shows the cumulative distribution of the margin \(\Delta P = P_1 - P_2\) for these objects, split by most-probable web type. The selected objects are overwhelmingly sheets and filaments, with 125,486  and 61,181  objects, respectively, together accounting for 83\% of the selection. For these two classes, the probability gap remains large even inside the entropy peak. The median margin is \(\Delta P \approx 0.45\) for sheets and \(\Delta P \approx 0.42\) for filaments, and fewer than 10\% of them have \(\Delta P<0.1\). Thus, for a typical sheet or filament in this entropy range, the dominant environment is still clearly identified even though the entropy lies near the peak of the global distribution.

This behaviour can be understood from the fact that the four web-type probabilities add up to one. An entropy of \(H\approx0.5\) can be reached in different ways. If only two types are involved, \(H=0.5\) is the maximum possible value for a two-class competition, \(\log 2/\log 4\), and it requires an almost even split; in this case the margin is forced to be small, for example, $(0.585, 0.415)$ gives
$H = 0.49$ and $\Delta P = 0.17$.
If a third type is involved, the same entropy can be compatible with a clearly dominant class, because part of the remaining probability is shared with another neighboring environment. 
For example,
$(0.75, 0.20, 0.05)$ gives $H = 0.50$ and $\Delta P = 0.55$. At fixed entropy, the margin therefore indicates how many environments are effectively competing for the object.

The two populations in \autoref{fig:margin_H05} correspond to these two regimes. Voids and knots lie at the extremes of the \( r\) range and have only one adjacent class. Therefore, when they become uncertain, the decision is close to a two-way contest with a sheet or a filament, and their margins are small, with medians of \(\Delta P \approx 0.17\) and \(0.22\), respectively. However, they are a minority in the entropy window, accounting for 17\% of the selection, and most voids and knots occur at much lower \(H\), where their classification is unambiguous. Sheets and filaments can instead leak probability to both of their neighboring environments; the third type increases the entropy while the dominant probability remains high, producing their larger margins. Finally, no object in the window exceeds \(\Delta P \approx 0.7\), which is the consistency limit imposed by the entropy selection itself, since a larger gap would force \(H\) below the selected range.

\subsection{Web-Type Fractions}
\label{sec:web_type_fraction}

\autoref{tab:count_fraction} reports the count fractions for both the object and random catalogs, averaged over 100 ASTRA realizations and 20 zones. For the random catalogs, the count fraction approximates the volume-filling fraction (VFF), since the randoms Poisson-sample the survey volume with the same footprint and redshift distribution as the data. These fractions average the labels of the individual realizations; counts based on the most-probable web type, as used in \autoref{sec:Classification_Uncertainty}, favor the intermediate classes and therefore give a somewhat larger share of sheets and filaments.

\begin{table}[htbp]
\centering
\vspace{10pt}
\small
\setlength{\tabcolsep}{6pt}
\renewcommand{\arraystretch}{1.15}
\begin{tabular}{cccccc}
\hline\hline
\multirow{2}{*}{Catalog} & \multirow{2}{*}{Tracer} & \multicolumn{4}{c}{Count fraction (\%)} \\
 & & Void & Sheet & Filament & Knot \\[2pt]
\hline
\multirow{4}{*}{\makebox[28pt][l]{Object}}
& BGS & $14.97 \pm 0.55$ & $33.25 \pm 1.52$ & $31.86 \pm 0.90$ & $19.92 \pm 1.99$ \\
& LRG & $18.37 \pm 0.52$ & $43.31 \pm 1.31$ & $28.69 \pm 0.86$ & $9.64 \pm 1.10$ \\
& ELG & $17.15 \pm 0.31$ & $53.04 \pm 0.77$ & $25.83 \pm 0.50$ & $3.99 \pm 0.30$ \\
& QSO & $18.77 \pm 0.39$ & $61.23 \pm 0.77$ & $18.60 \pm 0.62$ & $1.40 \pm 0.20$ \\
\hline
\multirow{4}{*}{\makebox[35pt][l]{Random}}
& BGS & $58.39 \pm 1.11$ & $29.87 \pm 1.08$ & $9.67 \pm 0.38$ & $2.07 \pm 0.19$ \\
& LRG & $46.79 \pm 1.15$ & $41.41 \pm 1.25$ & $10.59 \pm 0.33$ & $1.20 \pm 0.11$ \\
& ELG & $31.36 \pm 0.73$ & $52.86 \pm 0.76$ & $14.57 \pm 0.20$ & $1.21 \pm 0.05$ \\
& QSO & $22.92 \pm 1.02$ & $61.35 \pm 0.88$ & $14.90 \pm 0.36$ & $0.82 \pm 0.06$ \\
\hline
\end{tabular}
\caption{\small Count fractions by web type for each tracer, in both the object and random catalogs. Fractions are averaged over 100 ASTRA realizations and 20 zones; uncertainties are inter-zone standard deviations. For the random catalogs, these fractions estimate the volume-filling fractions (VFF).}
\label{tab:count_fraction}
\end{table}

In the object catalogs, sheets are the dominant class for all tracers, followed by filaments. For the random catalogs (VFF), the dominant environment varies: voids dominate for BGS and LRG, while sheets dominate for ELG and QSO, reflecting the different redshift ranges and number densities of each tracer.

The BGS VFF ($58\%$ void, $30\%$ sheet, $10\%$ filament, $2\%$ knot), by construction, is in very good agreement with the GAMA reference values at $4\,h^{-1}\,\mathrm{Mpc}$ smoothing ($59\%$, $29\%$, $10\%$, $1\%$; \citep{Eardley_2015}).
The galaxy count fractions for BGS ($15\%$ void, $33\%$ sheet, $32\%$ filament, $20\%$ knot) also agree reasonably with the GAMA values ($18\%$, $34\%$, $36\%$, $12\%$). 
This broad agreement in the galaxy count fraction validates the physical consistency of the classification and supports the interpretation of the fractions for the other tracers.

\subsubsection{Redshift Evolution of Web-Type Fractions}

\begin{figure}[htbp]
\begin{center}
\includegraphics[width=1\textwidth]{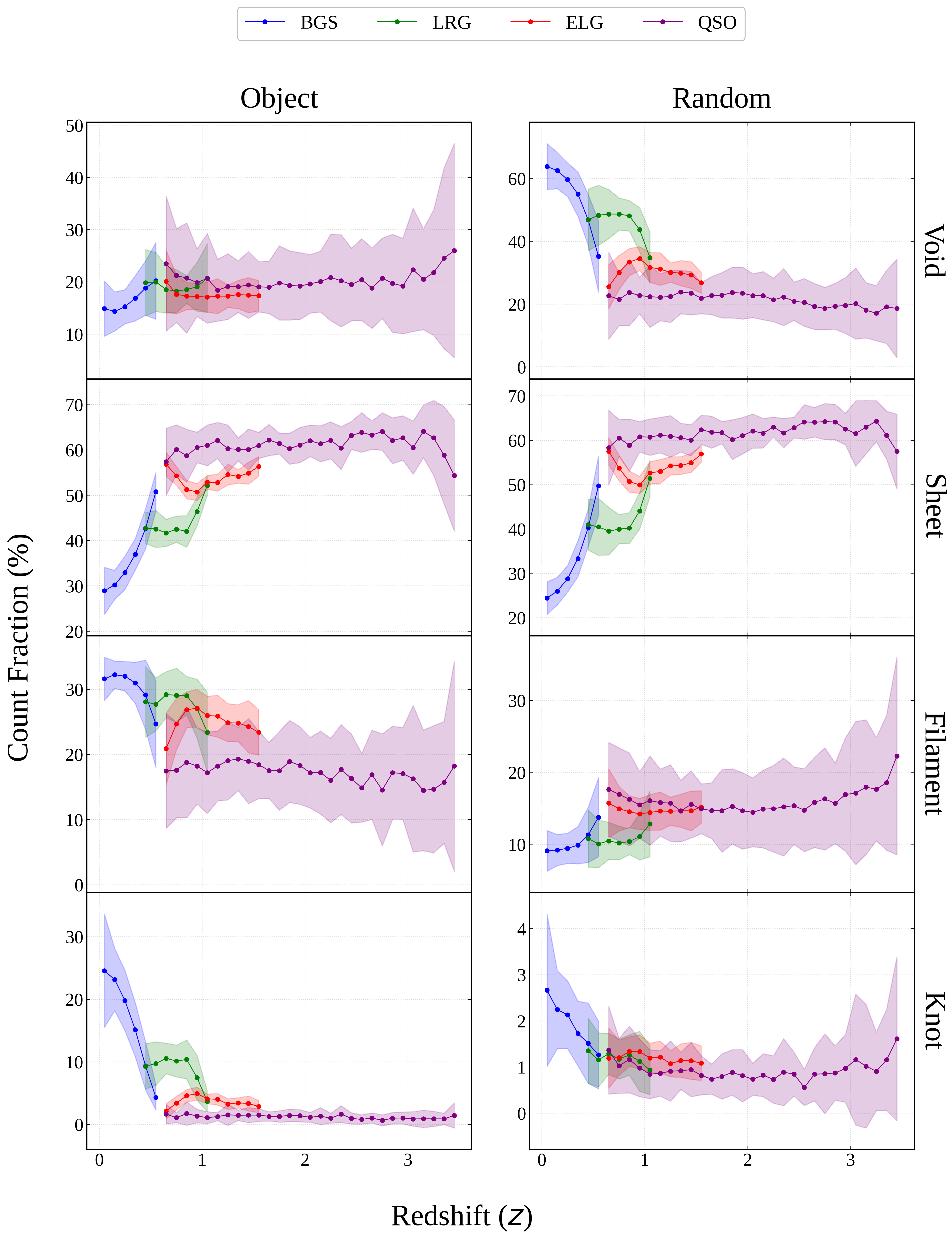}
\caption{\small Count fractions as a function of redshift for the object (left) and random (right) catalogs of each DESI tracer. Rows correspond to the four web types. Error bars show the inter-zone standard deviation.}
\label{fig:count_fraction}
\end{center}
\end{figure}

\autoref{fig:count_fraction} shows the count fractions as a function of redshift, computed in bins and averaged over the 20 zones.

LRG, ELG, and QSO show weak evolution in their fractions over their respective redshift ranges, both in the object and random catalogs. BGS shows the strongest evolution: in the random catalogs, the void VFF decreases from $63.8\%$ at low $z$ to $35.2\%$ at $z \sim 0.55$, while the sheet VFF rises from $24.4\%$ to $49.7\%$. In the object catalog, the sheet fraction increases from $28.9\%$ to $50.8\%$, the filament fraction drops from $31.6\%$ to $24.7\%$, and the knot fraction drops from $24.6\%$ to $4.3\%$. This behavior is consistent with BGS being flux-limited \citep{Ruiz_Macias_2020}: at higher redshift, the sample selects progressively more luminous galaxies, so the traced population is not equivalent across the full redshift range.

A complementary consistency check comes from the redshift bin around $z \sim 0.4$--$0.5$, where BGS and LRG both have objects and their fractions are in good agreement within the uncertainties. Since these two tracers are drawn from very different galaxy populations (BGS being a magnitude-limited sample of all galaxy types and LRG selecting massive, passively evolving galaxies) their convergence in web-type fractions at this redshift suggests that, at comparable number density and volume, the web-type fractions recovered by ASTRA depend little on the galaxy population used as tracer. This does not contradict the dependence of the fractions on number density and redshift range seen in \autoref{tab:count_fraction}: the fractions are set by how well the volume is sampled, not by the type of galaxy that samples it.

Similarly, in the overlap region $z \sim 0.9$--$1.1$, the LRG and ELG fractions agree within uncertainties for all environments \citep{valcin2025combinedtraceranalysisdesi} reinforcing the physical coherence of the ASTRA classification.

\subsection{Stellar Mass Distribution}
\label{sec:stellar_mass}

\begin{figure}[htbp]
\begin{center}
\includegraphics[width=0.9\textwidth]{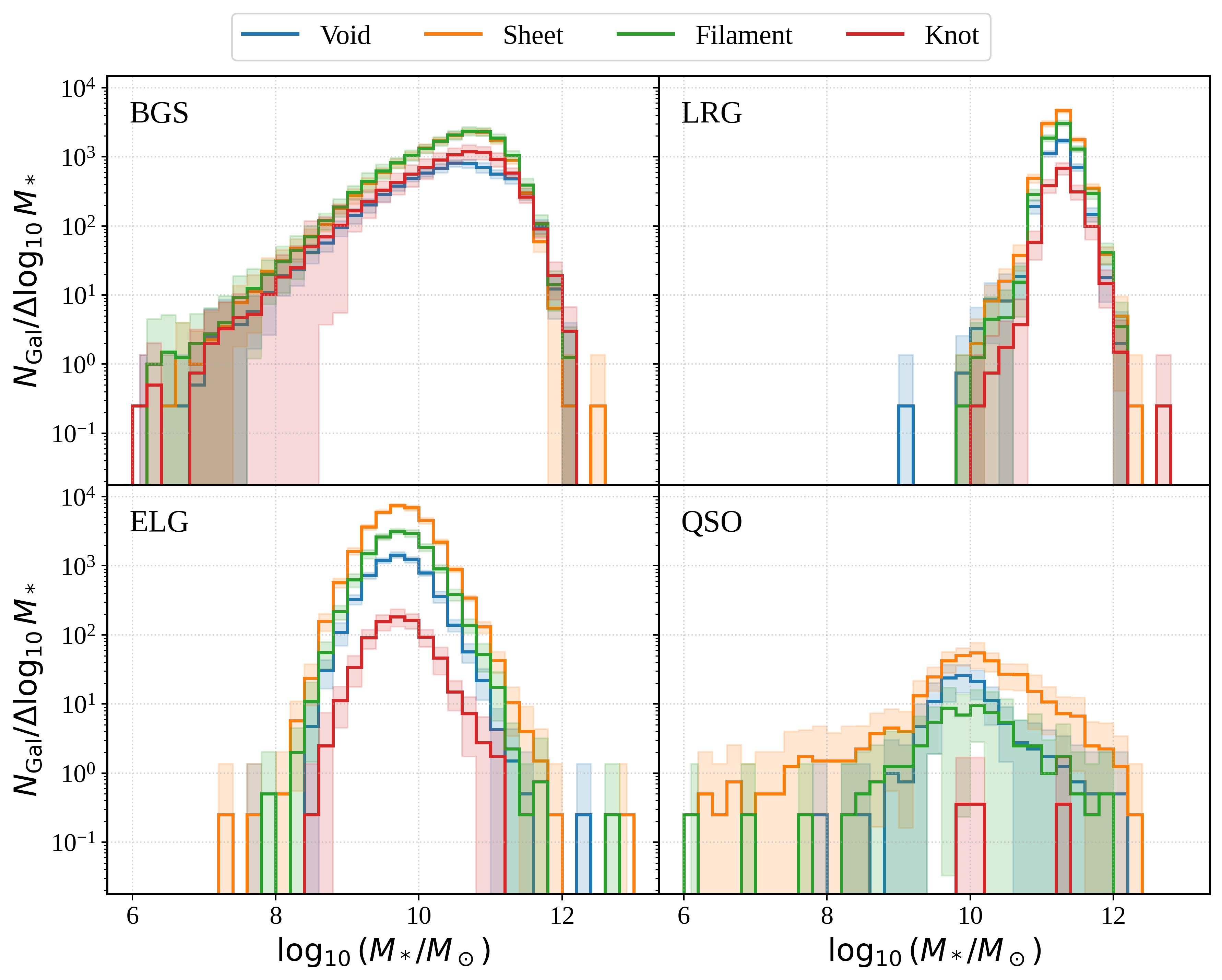}
\caption{\small Stellar mass distributions for the four tracers split by cosmic-web environment. Curves show counts per mass bin; shaded regions show the inter-zone dispersion.}
\label{fig:logM}
\end{center}
\end{figure}

\autoref{fig:logM} shows the stellar mass distribution $\log_{10}(M_\star/M_\odot)$ for each tracer split by web type. The distributions depend more strongly on tracer type than on environment: BGS is broad, LRG is shifted to higher masses with a narrow range, ELG is dominated by lower-mass systems, and QSO is limited by low number statistics. In all cases, sheets and filaments contain the most objects across the mass range, while knots are subdominant. This supports the picture that most stellar mass at these redshifts resides in intermediate-density environments.

For the BGS sample, the stellar mass fractions by environment ($14\%$ void, $33\%$ sheet, $35\%$ filament, $18\%$ knot; \autoref{tab:mass_fraction}) are broadly consistent with those reported for COSMOS \citep{Darvish_2017}, where the COSMOS \textit{field} category corresponds to our combined void and sheet ($45\%$ vs.\ $47\%$), filaments are comparable ($41\%$ vs.\ $35\%$), and clusters/knots are of similar order ($14\%$ vs.\ $18\%$).

\begin{table}[htbp]
\centering
\small
\textbf{Stellar mass fraction (\%)}\\
\vspace{1pt}
\rule{0.72\linewidth}{0.4pt}\\
\vspace{-10pt}
\rule{0.72\linewidth}{0.4pt}\\
\vspace{4pt}
\begin{tabular*}{0.72\linewidth}{@{\extracolsep{\fill}}cccc}
\multicolumn{4}{c}{\textbf{This work (BGS)}} \\
\hline
Void & Sheet & Filament & Knot \\
\hline
$13.90 \pm 0.55$ & $33.45 \pm 1.79$ & $34.59 \pm 1.26$ & $18.07 \pm 2.44$ \\
\hline
\end{tabular*}
\vspace{6pt}
\begin{tabular*}{0.72\linewidth}{@{\extracolsep{\fill}}ccc}
\multicolumn{3}{c}{\textbf{Darvish et al. \cite{Darvish_2017}}} \\
\hline
\hspace{15pt}Field & \hspace{-20pt}Filament & \hspace{-55pt}Cluster \\
\hline
\hspace{15pt}44.6 & \hspace{-20pt}41.1 & \hspace{-55pt}14.3 \\
\hline
\end{tabular*}
\caption{\small Stellar mass fraction by environment for BGS (this work) and COSMOS \citep{Darvish_2017}. The COSMOS \textit{field} corresponds to void + sheet in our classification. Uncertainties are inter-zone standard deviations.}
\label{tab:mass_fraction}
\end{table}

\subsection{Star Formation Rate and Environmental Segregation}
\label{sec:sfr}

A central question in galaxy evolution is how galaxy properties vary across large-scale cosmic-web environments, and what are the correlations between different cosmic web aspects (density, anisotropy, connectivity) with varying galaxy properties.
Observational studies have consistently found that galaxies in denser environments tend to be redder, more massive, and less actively star-forming than those in underdense regions \citep{Peng_2010, Darvish_2016, Parente_2024}. At the same time, void galaxies show higher star formation rates, bluer colors, and later-type morphologies than their counterparts in the field or in denser structures. The physical processes behind this segregation are still debated, and measuring them across a broad redshift range and with a well-defined classification scheme is key to making progress \citep{Peng_2010, Kovac_2013}.

As shown in \autoref{fig:density_vs_r}, the ASTRA rank parameter \(\langle r\rangle\) is not sharply correlated with the local number density. 
A single value of the local density is therefore compatible with very different values of \( r\). This distinction is important for the interpretation of the environmental trends below: the ASTRA classes cannot be interpreted as a pure local-density ranking.

This behaviour is consistent with previous work showing that geometric constructions on the galaxy distribution carry environmental information beyond the local density. 
  A close relative of the Delaunay tessellation used by ASTRA is the $\beta$-skeleton, a subgraph of the Delaunay tessellation that connects two galaxies whenever the lens-shaped region between them, whose width is set by the parameter $\beta$, contains no other galaxy. \citep{Su_rez_P_rez_2021} showed that six features of this graph, computed for $\beta=1$, predict the dark matter T-Web environment
  \citep{10.1111/j.1365-2966.2009.14885.x} with a global accuracy of
  \(74\%\). 
  In that study the most informative features were a graph-based estimate of the local density and the mean edge length to the first neighbours, but the contrast between the density at a galaxy and the average over its graph neighbours ranked next, showing that the arrangement of the neighbours adds information that the local density alone does not carry. 
  The $r$ parameter of ASTRA belongs to the same family of tessellation-based quantities, and the results of this section show again that it is not a simple local-density estimate.

Furthermore, a key methodological point in ASTRA is the role of the random catalogue. 
A density estimate built only from galaxies cannot trace underdense regions in a stable way, because by definition there are almost no galaxies there. 
In those regions the Delaunay tetrahedra around a galaxy become very large and few, so $n_i$ is set by the positions of a handful of distant neighbours and is dominated by noise; this is precisely where the scatter in \autoref{fig:density_vs_r} is largest. 
In ASTRA, instead, the empty space is populated by random points, and a void is identified by the dominance of randoms among the Delaunay neighbours of a galaxy. 
The quantity $r$ therefore remains well defined and informative in the low-density regime, where a purely galaxy-based density estimator becomes less effective. The behavior of $r$ near the edges of the footprint and in regions of reduced fiber-assignment completeness is quantified in Section \ref{sec:boundary_effects}.

In this context, the BGS sample offers a unique opportunity: it covers the low-redshift universe  with high spectroscopic completeness, and its overlap with GAMA and COSMOS allows direct comparisons with established results. We therefore restrict the star-formation analysis to BGS, which has the highest spectroscopic completeness and, being magnitude-limited, covers all galaxy types among the four DESI tracers. We apply a cut $-4 < \log(\mathrm{SFR}) < 3$ to remove the extreme tails of the distribution while preserving the bulk of the population.

\begin{figure}[htbp]
\begin{center}
\includegraphics[width=0.7\textwidth]{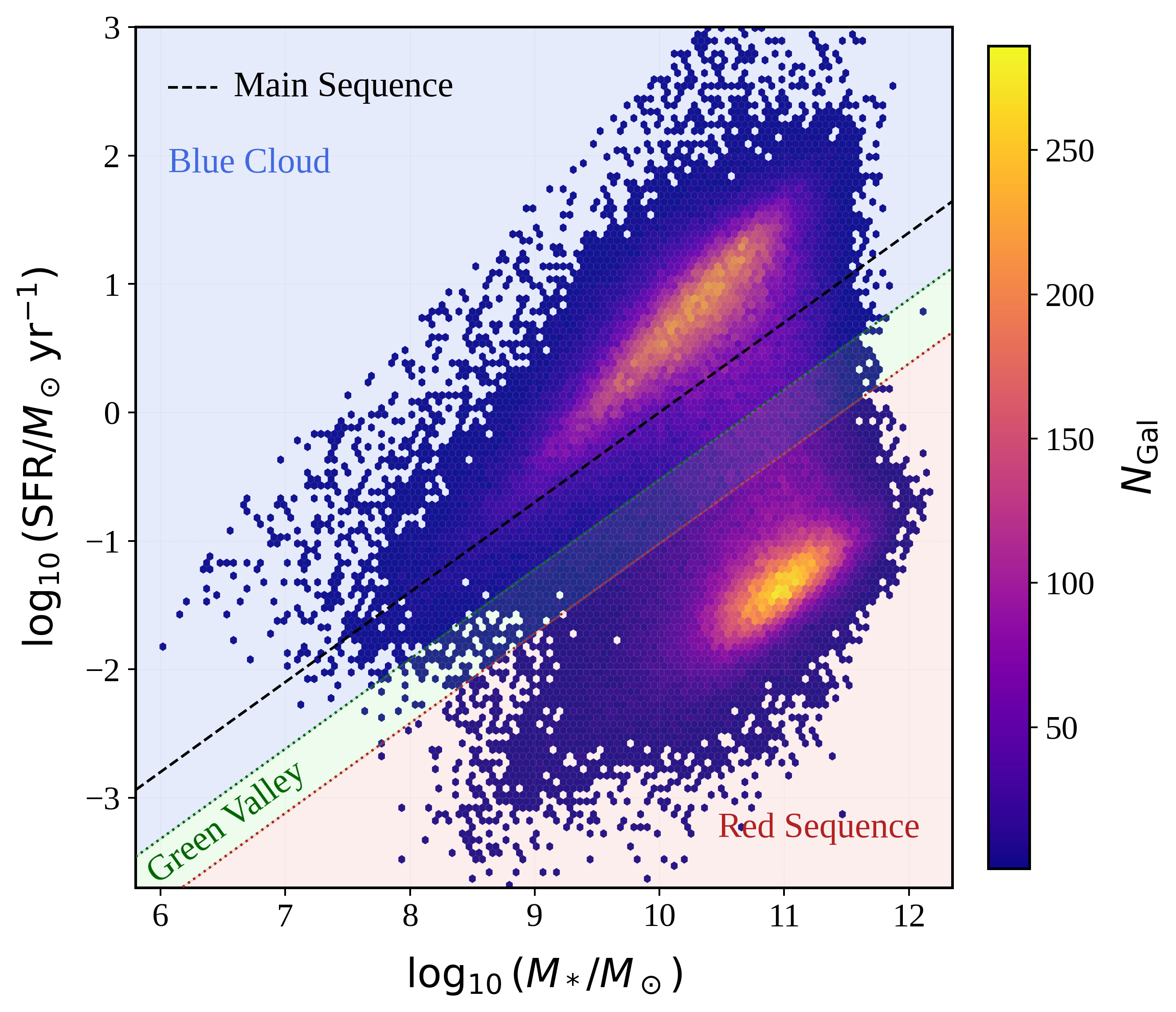}
\caption{\small $\log(M_\star/M_\odot)$ vs.\ $\log(\mathrm{SFR})$ for the BGS sample. The dashed line marks the star-forming main sequence. Shaded regions indicate the blue cloud, green valley, and red sequence. The dashed line is a visual guide, not a fit performed in this work.}
\label{fig:seq}
\end{center}
\end{figure}

\autoref{fig:seq} shows the $\log(M_\star)$--$\log(\mathrm{SFR})$ plane, using the physical quantities provided by the DESI EDR stellar-mass-emline VAC, for the BGS sample. The dashed line is shown only as a visual reference for the approximate location of the star-forming sequence. The distribution follows the star-forming main sequence \citep{Noeske_2007}, with the blue-cloud population above and the red sequence below. The green valley separates these two regimes \citep{Schawinski_2014, O_Kane_2024}, and the three populations are clearly visible in the BGS data, confirming that the sample covers the full range of star-formation activity needed for environmental studies.

\begin{figure}[htbp]
\begin{center}
\includegraphics[width=1\textwidth]{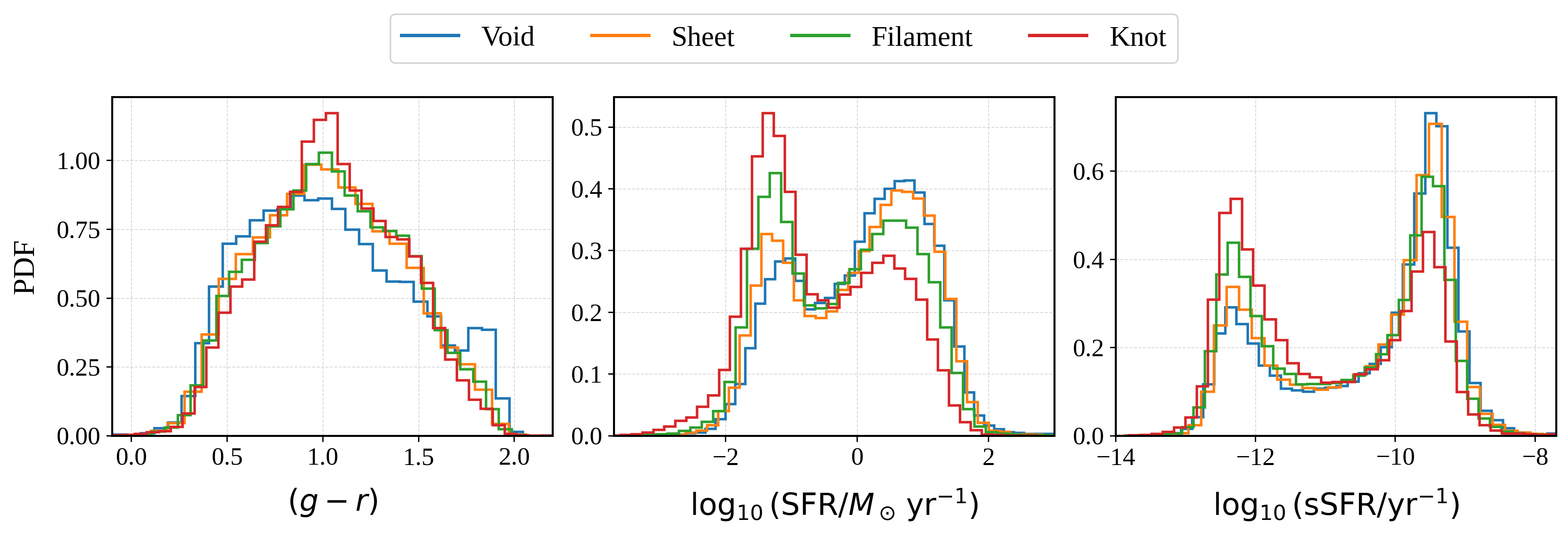}
\caption{\small Normalized distributions of $(g-r)$, $\log_{10}(\mathrm{SFR}/M_\odot\,\mathrm{yr}^{-1})$, and $\log_{10}(\mathrm{sSFR}/\mathrm{yr}^{-1})$ for BGS, split by cosmic-web environment.}
\label{fig:pdf_env}
\end{center}
\end{figure}

\autoref{fig:pdf_env} shows the normalized distributions of $(g-r)$ color, $\log(\mathrm{SFR})$, and $\log(\mathrm{sSFR})$ for each web type. The color distributions are broadly similar across environments, with knots being slightly redder and voids slightly bluer, but the difference is modest. The environmental signal becomes much clearer in the star-formation indicators: knots show the strongest contribution at low SFR and low sSFR, while voids contain the largest fraction of highly star-forming galaxies. Sheets and filaments lie in between. This pattern is consistent with the expected picture in which denser regions host preferentially quenched populations, while underdense environments preserve more actively star-forming systems.

\begin{figure}[htbp]
\begin{center}
\includegraphics[width=0.6\textwidth]{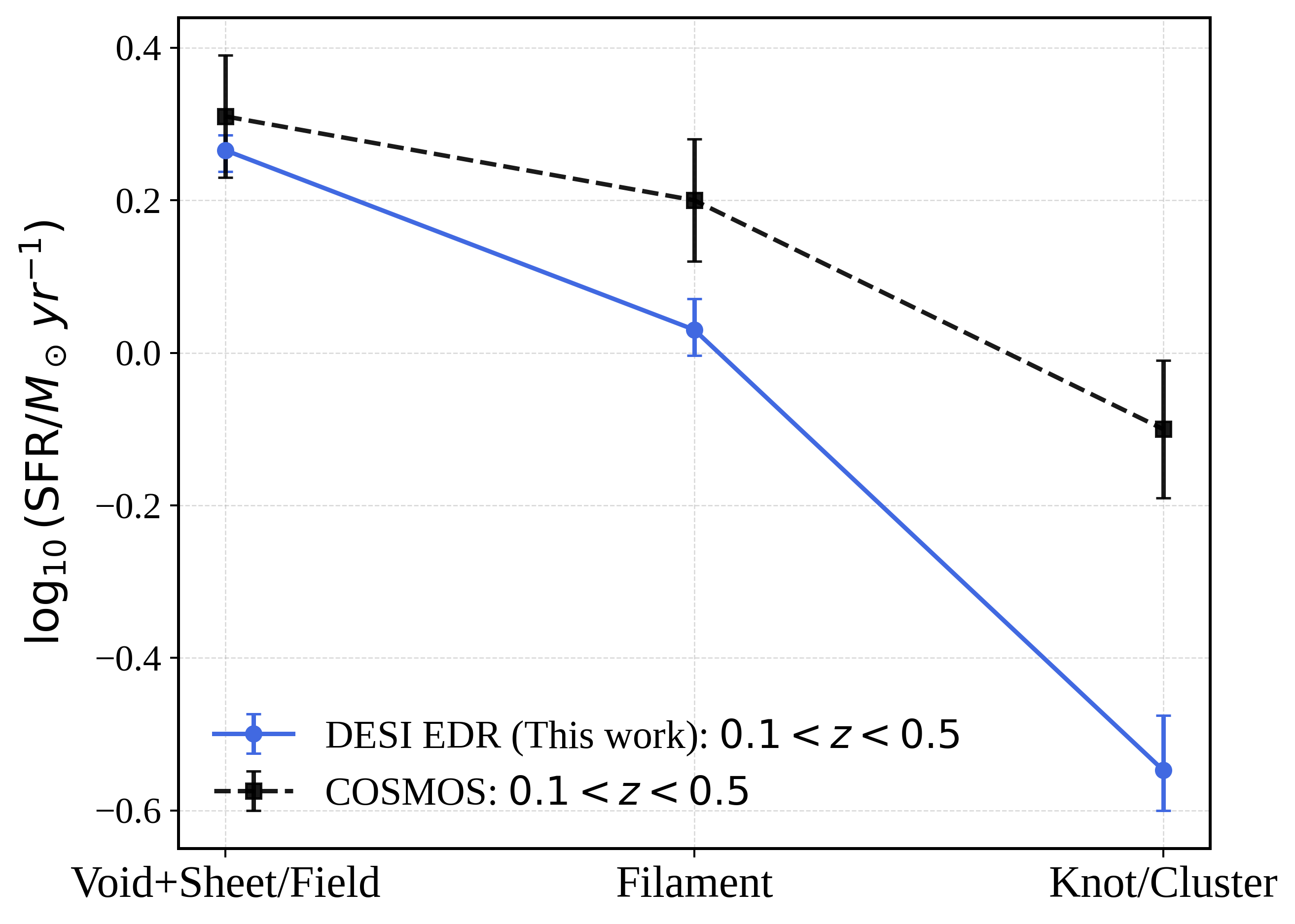}
\caption{\small Median $\log(\mathrm{SFR})$ as a function of environment for BGS galaxies in zone~0 (blue) compared with COSMOS \citep{Darvish_2017} (black). Both samples use $0.1 < z < 0.5$ and $\log(M/M_\odot) \geq 9.6$.}
\label{fig:sfr_z}
\end{center}
\end{figure}

To compare our results with a well-established observational reference in the same sky region, \autoref{fig:sfr_z} shows the median $\log(\mathrm{SFR})$ as a function of environment for BGS galaxies in zone~0, which spatially overlaps with COSMOS, alongside the results of \citet{Darvish_2017}. We apply the same redshift ($0.1 < z < 0.5$) and mass ($\log(M/M_\odot) \geq 9.6$) cuts used in their analysis. Both datasets show a clear and consistent gradient: SFR decreases systematically from voids and sheets toward knots. This agreement is notable because the two analyses differ in methodology. The COSMOS results use photometric redshifts and a 2D classification over $\sim 1.8\,\mathrm{deg}^2$ with the Multi-scale Morphology Filter applied to a two-dimensional density field, while our BGS results use spectroscopic redshifts and a full 3D ASTRA classification over $\sim 7\,\mathrm{deg}^2$ (the region within $1.5^\circ$ of the rosette center, see \autoref{sec:boundary_effects}) for the same zone, suggesting that the environmental gradient in SFR is robust to both the classification method and the data type.

\begin{figure}[htbp]
\begin{center}
\includegraphics[width=1\textwidth]{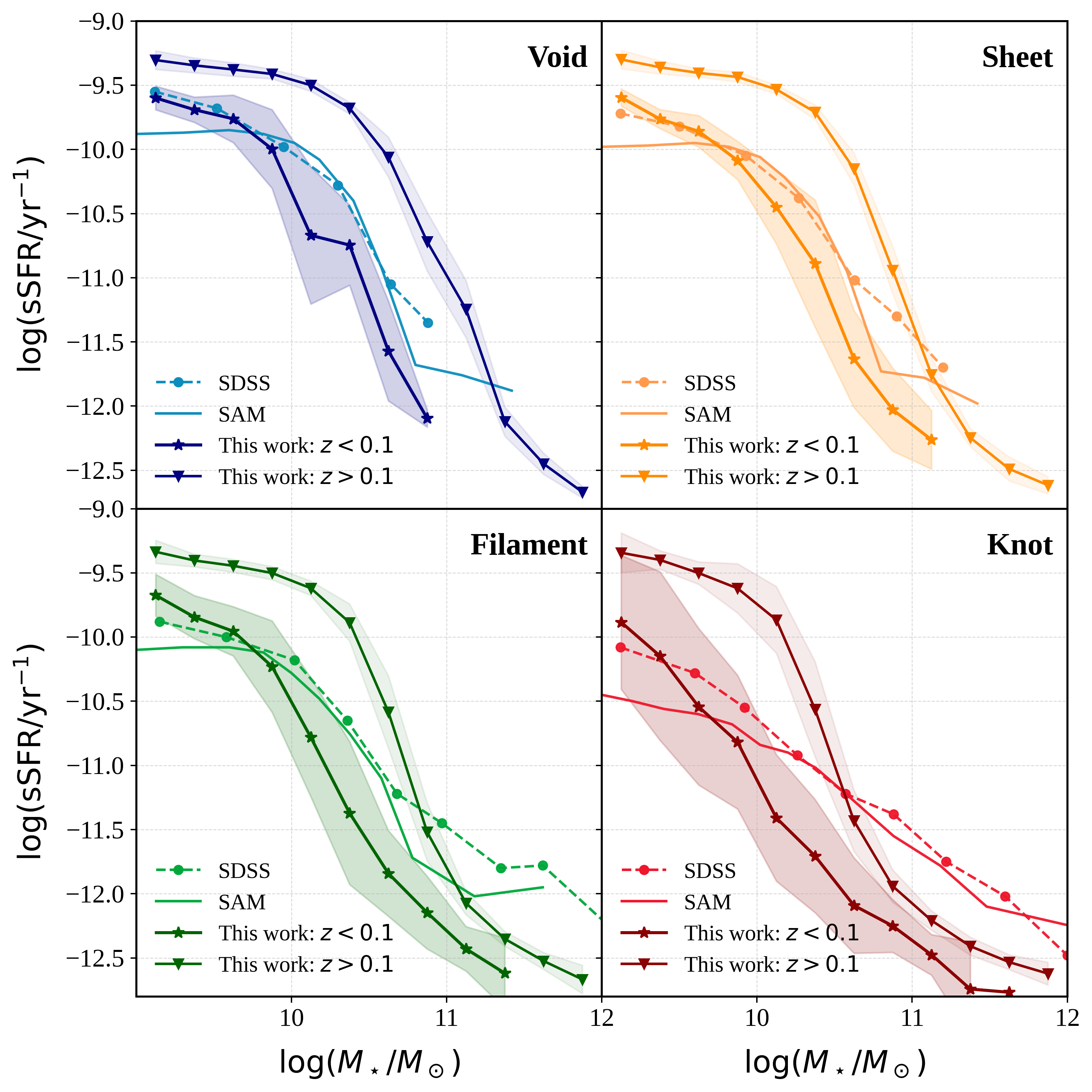}
\caption{\small Median $\log(\mathrm{sSFR})$ vs.\ stellar mass for BGS, split by environment and redshift range. Comparison data in \citep{Parente_2024} use SDSS DR16 and L-GALAXIES with the NEXUS+ web classification. Shaded bands show the inter-zone dispersion.}
\label{fig:ssfr_mass_env}
\end{center}
\end{figure}

A more detailed picture emerges from the sSFR--mass relation. \autoref{fig:ssfr_mass_env} shows the median $\log(\mathrm{sSFR})$ as a function of stellar mass for each web type, split into $z < 0.1$ and $z > 0.1$, and compared with SDSS observations and L-GALAXIES semi-analytic model with the NEXUS+ cosmic web finder \citep{Parente_2024}. In all environments, sSFR decreases with increasing stellar mass. At fixed mass, the environmental ordering is consistent across all panels: voids show the highest sSFR, knots the lowest, and sheets and filaments are intermediate. This ordering is stable across both redshift bins, suggesting that the connection between cosmic-web environment and star formation is already in place over the full BGS redshift range and is not a transient feature. The ASTRA-based environmental dependence is somewhat stronger than in the SDSS and L-GALAXIES references, particularly in filaments and knots, which may partly reflect differences in how the two classification schemes identify dense structures.

\subsubsection{Mutual Information Analysis}
\label{sec:nmi}

To quantify the statistical associations between galaxy properties as a function of environment, we compute the normalized mutual information (NMI) for each pair of variables in each web type. Variables are discretized into 30 bins after clipping the 1\% tails; uncertainties are estimated with a jackknife over 50 subsamples. The NMI is defined as:

\begin{equation}
\mathrm{NMI}(X,Y) = \frac{I(X;Y)}{\sqrt{H(X)\,H(Y)}},
\label{eq:nmi}
\end{equation}

where $I(X;Y)$ is the mutual information and $H(X)$, $H(Y)$ are the marginal entropies. $\mathrm{NMI} = 0$ indicates no association and larger values indicate stronger dependence.

\begin{figure}[htbp]
\begin{center}
\includegraphics[width=0.7\textwidth]{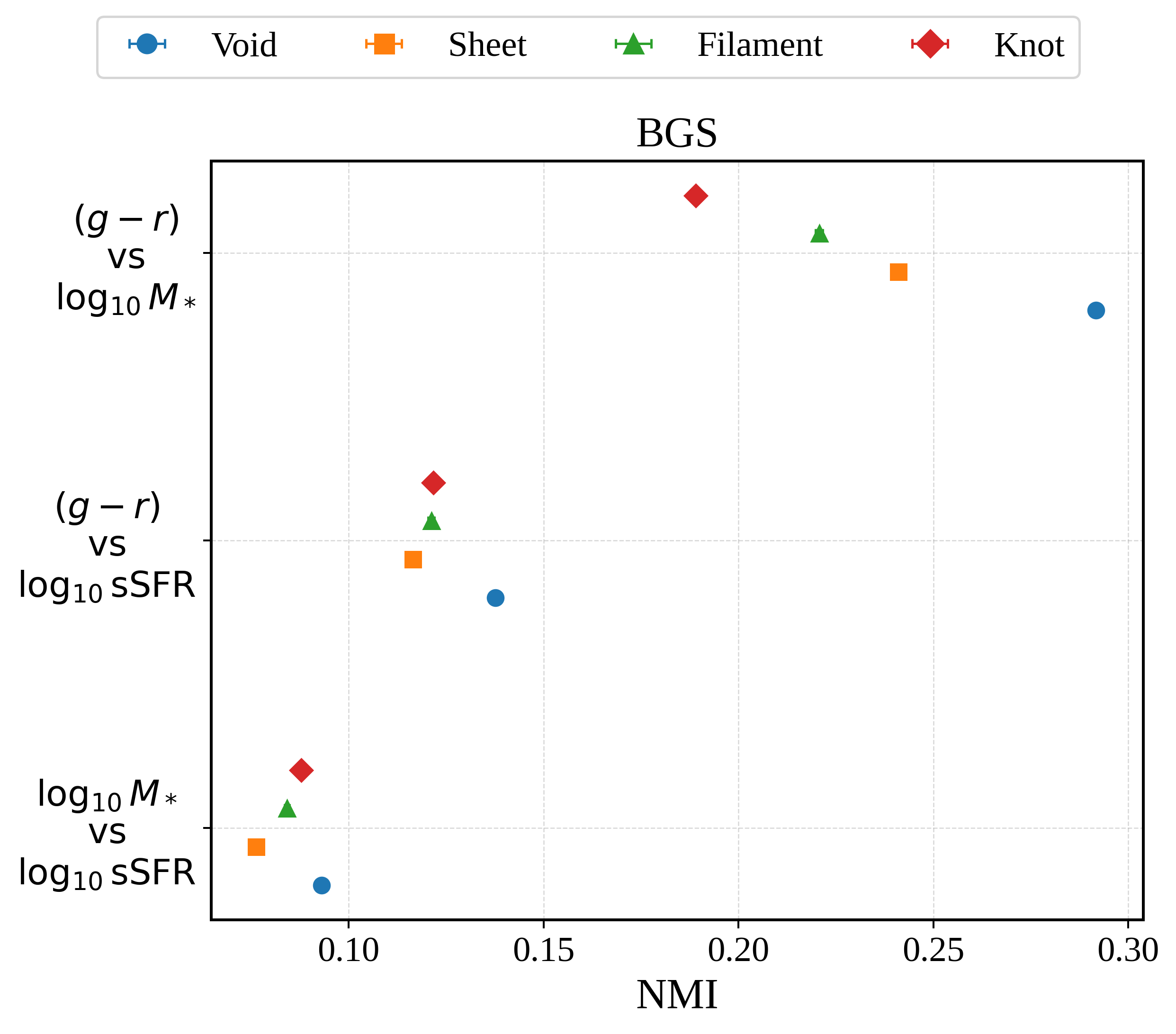}
\caption{\small Pairwise NMI between galaxy properties for the BGS sample, computed separately in each web type. Error bars show $1\sigma$ jackknife uncertainties.}
\label{fig:nmi}
\end{center}
\end{figure}

\begin{figure}[htbp]
\begin{center}
\includegraphics[width=0.8\textwidth]{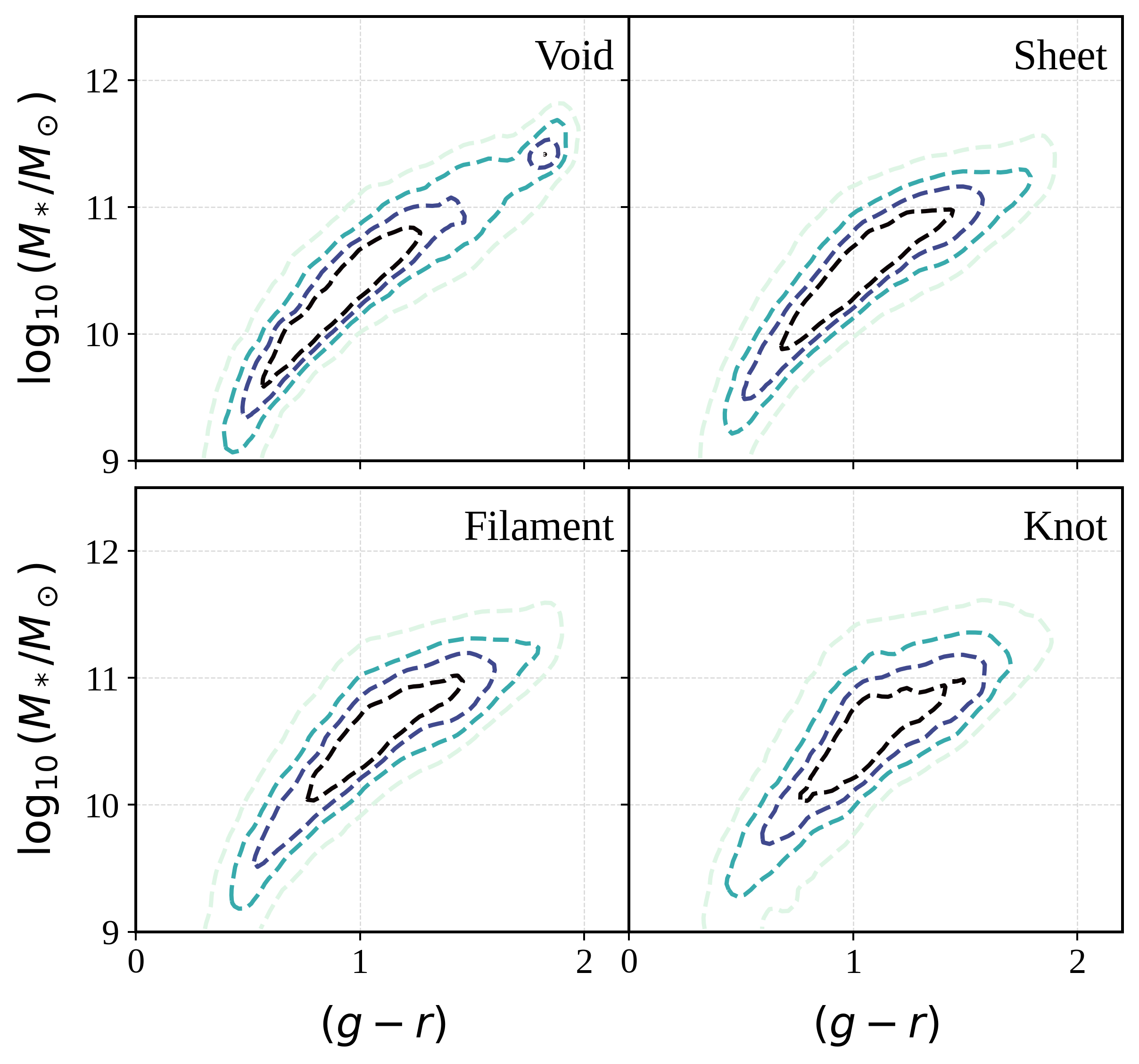}
\caption{\small $(g-r)$ vs.\ $\log(M_\ast/M_\odot)$ for the BGS sample in each web type. Contours show galaxy number density.}
\label{fig:color_mass}
\end{center}
\end{figure}

This analysis reveals a ranking of the NMI values that adds nuance to the sSFR--mass trends discussed above. 
Although the environmental ordering of the sSFR--mass relation is visually clear in \autoref{fig:ssfr_mass_env}, the NMI between $\log(M_\ast)$ and $\log(\mathrm{sSFR})$ is actually the weakest of the three pairs considered. 
The strongest association is between $(g-r)$ and $\log(M_\ast)$, followed by $(g-r)$ and $\log(\mathrm{sSFR})$. 
This tells us that optical color encodes more statistical information about both mass and star formation activity than mass and sSFR share with each other directly.

The color--mass NMI is highest in voids and decreases progressively toward knots. 
\autoref{fig:color_mass} shows why: in underdense regions, galaxies follow a tight, well-defined color--mass sequence with little scatter, while in denser environments the distribution broadens and the relation becomes less regular. 
This systematic decrease in NMI with density is a signature of the environment actively dispersing the color--mass relation, likely through the mix of physical processes that operate in dense regions (quenching, mergers, tidal interactions) and that add scatter beyond what stellar mass alone would predict \citep{Peng_2010, Kovac_2013}.

The color--sSFR and mass--sSFR pairs tell a different story: their NMI is highest in voids, reaches a minimum in sheets, and rises again from sheets through filaments to knots, without recovering the void value. 
This non-monotonic pattern, with a minimum at sheets, suggests that the way star-formation activity connects to mass and color is not a simple continuous function of density. 
In voids, and to a lesser extent in knots, galaxies tend to be sorted into well-separated star-forming and quenched sequences, which makes the pairwise statistical associations sharper. 
In sheets, the two sequences overlap more, the population mix is less clean, and the associations weaken; filaments are intermediate between sheets and knots. 
A possible reading of this is that sheets are transition zones where galaxies are in the process of moving from one sequence to the other, so neither the star-forming nor the quenched population fully dominates, and that this mixing is progressively resolved as galaxies move into filaments and knots. 
In this picture, the large-scale environment does not simply shift the mean star-formation activity; it regulates the pace at which galaxies move between states, with sheets playing a central role in that transition.

\section{Discussion and Future Work}
\label{sec:discussion}

\subsection{Validation and Physical Consistency}

The ASTRA classification of the DESI EDR recovers physically consistent results across all four tracers. The BGS volume-filling fractions reproduce GAMA by construction; the independent validation of the adopted thresholds comes from the galaxy-count fractions, which also agree with GAMA, and from the stellar mass and star formation results discussed below. The star formation results for BGS are consistent with COSMOS \citep{Darvish_2017} and with SDSS-based results at $z < 0.1$ \citep{Parente_2024}: in all cases, denser environments host more quenched populations at fixed stellar mass, and the environmental ordering of the sSFR--mass relation is stable.

The ASTRA approach offers a specific advantage in this context: by simultaneously classifying all four environments (voids, sheets, filaments, and knots), it avoids the asymmetry of methods that focus only on one or two components. The inclusion of voids is particularly relevant, since underdense regions are often omitted from environment catalogs despite being the dominant volume component for the BGS and LRG samples (\autoref{tab:count_fraction}).

\subsection{New Results for BGS}

The NMI analysis provides results that go beyond the low-redshift validation. 
The non-monotonic behavior of the color--sSFR and mass--sSFR associations across 
environments, with a minimum in sheets, suggests that the effect of large-scale 
environment on star-formation-related quantities is not well described as a simple 
continuous trend with density. 
Instead, the pattern is more consistent with a change in the 
relative fractions of star-forming and quenched galaxies, rather than a 
modification of the star-forming sequence itself \citep{Darvish_2016, Peng_2010}. 
On the other hand, the color--mass association does follow a monotonic trend with 
environment, from a tighter relation in voids to a broader scatter in knots. Put together, these results provide a new observational baseline for 
testing galaxy evolution models in a DESI context.

\subsection{Redshift-space Distortions and Environment-dependent Clustering}

Since the ASTRA classification is performed in redshift space, redshift-space distortions can affect the apparent geometry of the identified environments. In particular, Fingers of God elongate the densest knots along the line of sight. This effect can convert compact knots into filament-like structures pointing towards the observer. On large scales, the Kaiser effect compresses structures along the line of sight and enhances their apparent contrast. The random catalog shares the angular and radial selection of the data, but it does not contain peculiar velocities, so these distortions should have a different signature in the random data.

\begin{figure}[htbp]
\begin{center}
\includegraphics[width=0.98\textwidth]{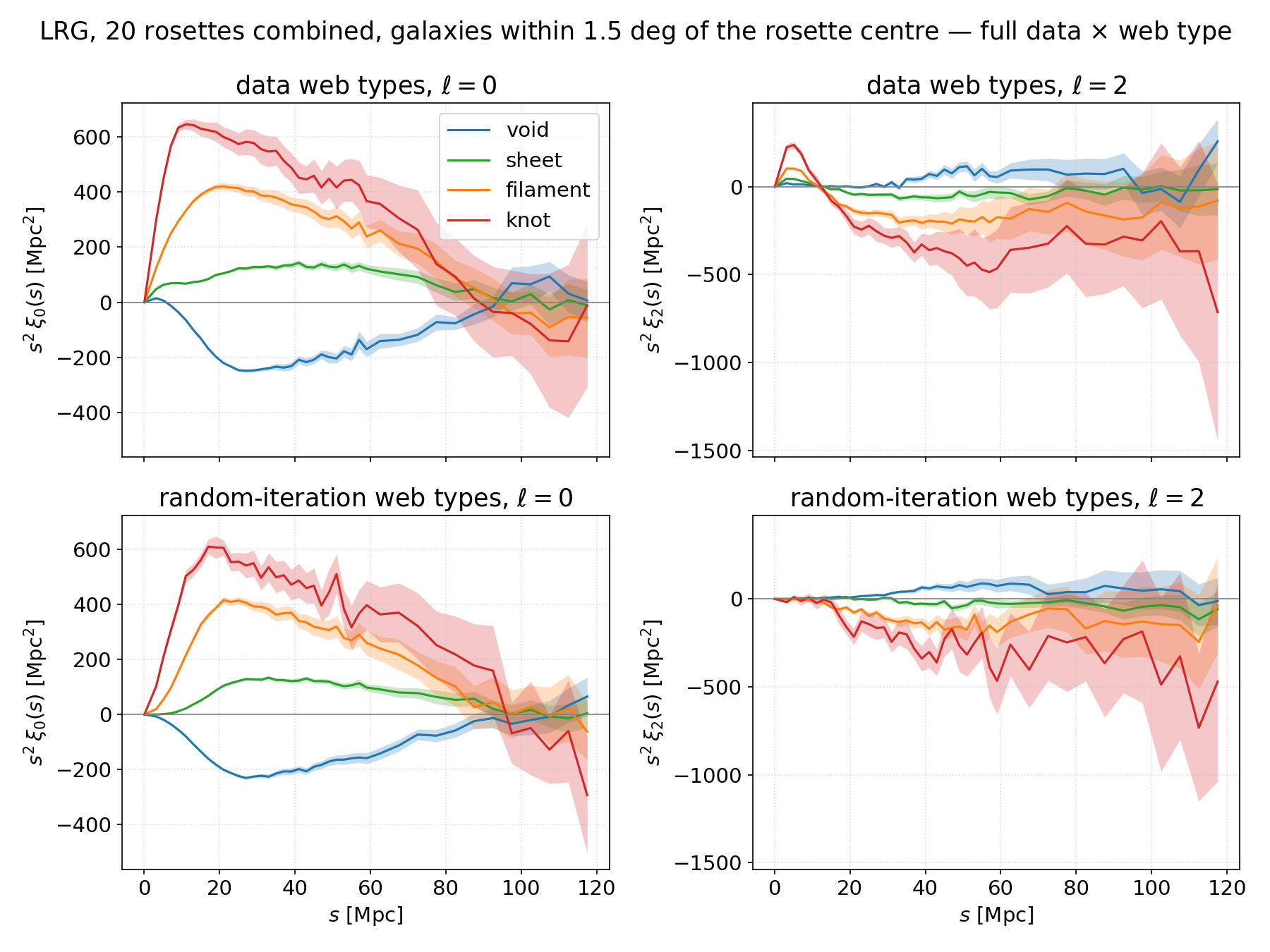}
\caption{\small Monopole (left column) and quadrupole (right column) of the cross-correlation function between the full LRG sample and subsets split by web type, for the 20 rosettes combined and restricted to $\theta \leq 1.5^\circ$. Top row: the subsets are LRG galaxies, split by their most-probable web type. Bottom row: the subsets are the random points of a single ASTRA realization, split by the web type assigned to each random point in that realization. The shaded bands show leave-one-rosette-out jackknife uncertainties. The separated monopoles and the distinct, non-zero quadrupoles show that the redshift-space clustering depends on the environment; the positive small-scale quadrupole of knots and filaments is present only when the subsets are galaxies.}
\label{fig:multipoles}
\end{center}
\end{figure}

In this work, redshift-space distortions are treated as a signal rather than only as a systematic. The ASTRA environment information can be used to compute the monopole and quadrupole of the cross-correlation function between the full galaxy sample and the galaxies split by environment. As an illustration, \autoref{fig:multipoles} shows this measurement for the LRG sample. The four environments show clearly separated monopoles and distinct, non-zero quadrupoles, indicating that the anisotropic pattern produced by redshift-space distortions depends on environment. The bottom panels show that a signal is also recovered when the subsets are not galaxies but the random points of a single ASTRA realization, split by the web type assigned to each random point in that realization. The random points sample the same environments as the galaxies but carry no peculiar velocities. Of special relevance is the fact that the quadrupole of knots and filaments differs between the galaxy subsets and the random-point subsets at scales below \(20\,{\rm Mpc}\): the positive small-scale quadrupole appears only when the subsets are galaxies, showing the impact of peculiar velocities on small scales.

This is a central added value of the catalog. Keeping the redshift-space distortions as a function of environment, instead of treating them only as a systematic to be removed, provides a valuable tool for the community working on the link between the dark matter distribution and the galaxy distribution. 
For instance, in Halo Occupation Distribution (HOD) parameterizations, the environmental dependence of the clustering multipoles could help constrain how galaxies populate their halos \citep{Prada_2025}. 
It also opens the possibility of using these multipoles to constrain cosmological parameters, which will be explored in a follow-up paper.

\subsection{Future Work}

Several extensions of this work are natural. The most direct is to apply ASTRA to the DESI DR1 catalogs \citep{desicollaboration2025datarelease1dark}, which cover much larger volumes and contain significantly more objects. 
The probabilistic outputs of ASTRA also open specific opportunities that are not explored here. The NMI analysis presented in \autoref{sec:nmi} suggests that sheets are the environment where the statistical associations between galaxy properties are weakest, suggesting that sheets play an active role in moving galaxies between star-forming and quenched states. 

The full set of membership probabilities can be used to define cleaner environment subsamples with stricter confidence cuts, to isolate galaxies in genuine transition regions, or to construct weighted statistics that propagate classification uncertainty directly into environmental measurements. 
These tools are particularly well suited to testing whether the weakening of the color--sSFR and mass--sSFR associations in sheets reflects a real physical process or a classification artifact.

On the galaxy property side, future analyses can extend well beyond what is explored here by combining the ASTRA environment catalog with other EDR value-added catalogs. 

\begin{itemize}

    \item 
The \textsc{FastSpecFit} Spectral Synthesis and Emission-Line Catalog provides homogeneous measurements of emission-line fluxes, equivalent widths, and continuum properties for all DESI tracers, enabling environmental studies of ionization state, gas-phase metallicity, and specific star formation activity beyond the SED-based estimates used in this work. 

    \item 
The \textsc{PROVABGS} (PRObabilistic Value-Added BGS) Catalog offers Bayesian physical parameter estimates (including star formation histories, stellar ages, and metallicities) for the BGS sample, which are directly relevant to understanding why the color--mass relation tightens in voids and broadens in knots. 

    \item 
For the AGN population, the AGN Host Galaxy Physical Properties Catalog and the Broad Absorption Line Quasar Catalog provide the necessary information to study how black hole activity depends on large-scale environment and whether it contributes to quenching in denser regions. 

    \item 
For the QSO tracer, the BAL and Mg~\textsc{ii} Absorber Catalogs open the possibility of studying the connection between absorber properties and cosmic-web environment along the line of sight. 

    \item 
The Uchuu--DESI Catalog, which provides a simulation-based mock counterpart to the DESI observations, offers a natural framework for comparing the observed environmental trends with theoretical predictions.
\end{itemize}

Finally, the public catalog released with this work (\autoref{sec:open_data}) provides a direct input for cosmological applications, including void statistics, environment-dependent clustering, and tests of the halo--environment connection. These applications will become more powerful with the larger DESI data releases and represent a natural next step for incorporating cosmic-web information into DESI's broader scientific program.

\section{Open data and code availability}
\label{sec:open_data}

The full analysis pipeline used in this work is publicly available as an open-source repository at
\url{https://github.com/forero/ASTRA-DESI}. The exact version of the code used for this paper is frozen in the tagged release \texttt{edr-paper-v1}\footnote{\url{https://github.com/forero/ASTRA-DESI/releases/tag/edr-paper-v1}}.

The resulting ASTRA--DESI EDR data products are archived on Zenodo under the DOI
\href{https://zenodo.org/records/22694998}{10.5281/zenodo.22694998}.

A detailed description of the released directory layout, file naming conventions, and column definitions is provided in \autoref{app:data_release}.

\section{Conclusions}
\label{sec:conclusions}

We present the first public cosmic-web environment catalog built on 
any DESI data release. Using \textsc{ASTRA} 
\citep{foreroromero2025cosmicwebclassificationstochastic}, we assign 
each object in the DESI EDR to a void, sheet, filament, or knot 
environment across 20 rosette fields and four extragalactic tracers 
(BGS, LRG, ELG, QSO), with per-object membership probabilities 
derived from 100 stochastic realizations. All data products and the 
analysis pipeline are publicly available at 
\href{https://zenodo.org/records/22694998}{doi:10.5281/zenodo.22694998}.

\paragraph{A new resource for the community.}

Having all four DESI EDR tracers classified under a single, 
self-consistent framework, covering a wider redshift range than 
any individual reference survey, is the primary new capability this 
catalog provides. The web-type fractions reported in 
\autoref{sec:web_type_fraction}, which span from BGS at $z \sim 0.1$ 
to QSO at $z \sim 3$, have not been reported before in the literature 
for DESI. The full probabilistic outputs (membership probabilities and 
classification entropies, \autoref{sec:Classification_Uncertainty}) 
are also released, making it possible for future users to apply 
confidence cuts or propagate classification uncertainty into their 
own analyses.

\paragraph{The BGS classification is physically reliable.}

The results in \autoref{sec:web_type_fraction}--\ref{sec:sfr} show 
that the \textsc{ASTRA} classification of BGS is consistent with every 
benchmark we could apply: volume-filling fractions match GAMA 
\citep{Eardley_2015}, stellar mass fractions are broadly consistent with 
COSMOS \citep{Darvish_2017}, and the sSFR--mass environmental ordering 
matches SDSS \citep{Parente_2024}. What makes this agreement 
meaningful is not the match itself (the VFF calibration was designed 
to reproduce GAMA by construction) but that the same calibration 
simultaneously recovers the correct galaxy-count fractions, stellar 
mass fractions, and star formation gradients, none of which were 
used in the calibration. This consistency across independent 
observables confirms that the thresholds in 
\autoref{tab:classification_adopted} capture real physical structure.

\paragraph{The NMI analysis opens new possibilities.}

The mutual information results in \autoref{sec:nmi} show that 
across all environments, the strongest pairwise association is between 
$(g-r)$ color and stellar mass, and this association grows 
progressively stronger from knots to voids.
In the associations involving the sSFR the NMI is highest in voids, 
reaches a minimum in sheets, and rises again through filaments to knots, 
a non-monotonic pattern that cannot be explained by a 
simple density-quenching picture. Together, these two trends suggest 
that the color--mass association, rather than the color distribution 
itself, is the quantity most sensitive to large-scale environment: the 
cosmic web shapes the color--mass relation gradually and continuously, 
while its effect on the associations involving star formation activity 
is strongest in voids and weakest in sheets. 
These measurements could become a concrete target for galaxy evolution models to reproduce. 

\paragraph{Outlook.}

The most immediate extension is to apply \textsc{ASTRA} to the DESI 
DR1 \citep{desicollaboration2025datarelease1dark}, 
where the much larger volume will allow the NMI patterns found here 
to be tracked as a function of redshift.
Another opportunity beyond sample size is to combine the 
environment catalog with the DESI value-added catalogs: spectral 
diagnostics and Bayesian stellar histories will extend the analysis 
beyond the SED-based quantities used here, AGN catalogs will test 
whether black hole activity drives quenching in knots, and 
simulation-based products such as Uchuu--DESI will allow a direct 
comparison with halo occupation models. Together, these extensions 
will move the analysis from describing environmental trends to 
testing the physical mechanisms behind them.

\appendix
\section{Description of the data package}
\label{app:data_release}

The dataset associated with this work is available on Zenodo
(\href{https://doi.org/10.5281/zenodo.22694998}{doi:10.5281/zenodo.22694998})
and is distributed in three files: \texttt{raw.tar.gz}, \texttt{classification.tar.gz} and \texttt{probabilities.tar.gz}. 
After uncompressing the files, each one generates a folder.
Below we describe, folder by folder, the organization and the fields of the tables.

\subsection{\texttt{raw.tar.gz}: merged catalogs (real + random data) per zone.}
\label{app:create_raw}

This folder contains 20 compressed FITS files (\texttt{.fits.gz}), one per rosette zone:
\[
\texttt{zone\_00.fits.gz},\ \texttt{zone\_01.fits.gz},\ \ldots,\ \texttt{zone\_19.fits.gz}.
\]
Each file integrates, in a single table, the \emph{real} objects in the zone and the 100 realizations of the associated \emph{random} catalog. The identification of the scientific tracer and the origin (real vs.\ random) is performed via the \texttt{TRACERTYPE} and \texttt{RANDITER} columns, respectively.

The stellar mass and star formation rate (SFR) included in this dataset are taken from the DESI Early Data Release (EDR) value-added catalogs \texttt{stellar-mass-emline} VAC. These quantities are derived through spectral energy distribution (SED) fitting using the CIGALE code, following the methodology described in Zou et al.\ \cite{Zou_2024}. The catalog provides homogeneous estimates of stellar mass and SFR for DESI galaxies based on multi-band photometry and emission-line information. We use the publicly available release and directly adopt the \texttt{SED\_MASS} and \texttt{SED\_SFR} columns for all tracers.

\paragraph{Meaning of the columns.}
\begin{itemize}
\item \textbf{\texttt{TARGETID}} (\texttt{int64}): unique identifier inherited from the EDR for each object.
\item \textbf{\texttt{RA}}, \textbf{\texttt{DEC}} (\texttt{float64}, deg): right ascension and declination in ICRS (degrees), as reported in the EDR.
\item \textbf{\texttt{Z}} (\texttt{float64}): dimensionless redshift from the EDR.
\item \textbf{\texttt{ROSETTE\_R}} (\texttt{float64}, deg): angular distance from the object to the center of the corresponding EDR rosette, in degrees.
\item \textbf{\texttt{XCART}}, \textbf{\texttt{YCART}}, \textbf{\texttt{ZCART}} (\texttt{float64}, comoving Mpc): comoving Cartesian coordinates calculated with the Planck 2018 cosmology adopted in this work.
\item \textbf{\texttt{TRACERTYPE}} (\texttt{string}): tracer type; values used: \texttt{BGS\_ANY}, \texttt{LRG}, \texttt{ELG}, \texttt{QSO}.
\item \textbf{\texttt{RANDITER}} (\texttt{int32}): iteration index of the random catalog. By convention, \texttt{-1} indicates a \emph{real} data object; \texttt{0}–\texttt{99} indicate the originating random realization.
\item \textbf{\texttt{SED\_SFR}} (\texttt{float64}): star formation rate (SFR) derived from spectral energy distribution (SED) fitting, based on the DESI EDR value-added catalogs.
\item \textbf{\texttt{SED\_MASS}} (\texttt{float64}): stellar mass estimated from SED fitting.
\item \textbf{\texttt{FLUX\_G}}, \textbf{\texttt{FLUX\_R}} (\texttt{float64}): observed fluxes in the $g$ and $r$ photometric bands, respectively.
\end{itemize}

\autoref{tab:dict_raw} summarizes the column dictionary applicable to all \texttt{zone\_xx.fits.gz} files in this folder.

\begin{table}[h]
\centering
\caption{Column dictionary for \texttt{zone\_xx.fits.gz} in \texttt{raw.tar.gz}.}
\label{tab:dict_raw}
\begin{tabular}{llll}
\cline{1-4}
\textbf{Name} &
\textbf{Type} &
\textbf{Units} &
\textbf{Description} \\
\cline{1-4}
\texttt{TARGETID}   & int64   & --  & Unique identifier inherited from EDR. \\
\texttt{RA}         & float64 & deg & Right ascension (ICRS). \\
\texttt{DEC}        & float64 & deg & Declination (ICRS). \\
\texttt{Z}          & float64 & --  & Redshift. \\
\texttt{ROSETTE\_R}      & float64 & deg & Angular distance from the object to the center of the rosette \\
\texttt{XCART}      & float64 & Mpc & Comoving Cartesian coordinate $x$. \\
\texttt{YCART}      & float64 & Mpc & Comoving Cartesian coordinate $y$. \\
\texttt{ZCART}      & float64 & Mpc & Comoving Cartesian coordinate $z$. \\
\texttt{TRACERTYPE} & string  & --  & Tracer (\texttt{BGS\_ANY}, \texttt{LRG}, \texttt{ELG}, \texttt{QSO}). \\
\texttt{RANDITER}   & int32   & --  & Record origin: \texttt{-1} real data; \texttt{0}--\texttt{99} random realization. \\
\texttt{SED\_SFR}   & float64 & $M_\odot\,\mathrm{yr}^{-1}$ & Star formation rate from SED fitting. \\
\texttt{SED\_MASS}  & float64 & $M_\odot$ & Stellar mass from SED fitting. \\
\texttt{FLUX\_G}    & float64 & nanomaggies & Observed flux in the $g$ band. \\
\texttt{FLUX\_R}    & float64 & nanomaggies & Observed flux in the $r$ band. \\
\cline{1-4}
\end{tabular}
\end{table}

\subsection{\texttt{classification.tar.gz}: neighbor statistics for web-type classification}
\label{app:class}

This folder contains 20 compressed FITS files (\texttt{.fits.gz}), one per rosette zone, following the same naming convention described in the previous sections:
\texttt{zone\_00\_classified.fits.gz}, \texttt{zone\_01\_classified.fits.gz}, \ldots, \texttt{zone\_19\_classified.fits.gz}.

Each row corresponds to a central object identified by \texttt{TARGETID}, evaluated against a specific random realization indexed by \texttt{RANDITER} $\in [0,99]$.

To construct these files, the complete dataset for each zone (real and random) is considered, and the local Delaunay neighborhood of each central \texttt{TARGETID} is evaluated separately for every random realization. Consequently, the same real \texttt{TARGETID} appears up to 100 times (once per random realization), since its neighbor configuration changes depending on the associated random catalog.

The distinction between real and random central objects is therefore encoded exclusively in the boolean column \texttt{ISDATA}.

\autoref{tab:dict_classified} summarizes the column dictionary applicable to all \texttt{zone\_xx\_classified.fits.gz} files in this folder.

\begin{table}[h]
\centering
\small
\caption{Column dictionary for \texttt{zone\_xx\_classified.fits.gz} in \texttt{classification.tar.gz}.}
\label{tab:dict_classified}
\begin{tabular}{lll}
\cline{1-3}
\textbf{Name} & \textbf{Type} & \textbf{Description} \\
\cline{1-3}
\texttt{TARGETID}   & int64  & Central object evaluated. \\
\texttt{RANDITER}   & int32  & Index of the random realization used to evaluate the local neighborhood (\texttt{0}--\texttt{99}). \\
\texttt{ISDATA}     & bool   & Origin of the central object; \texttt{True} for real objects, \texttt{False} for random objects. \\
\texttt{NDATA}      & int32  & Number of neighbors that are real-data objects in that evaluation. \\
\texttt{NRAND}      & int32  & Number of neighbors that are random-catalog objects in that evaluation. \\
\texttt{TRACER\_ID}  & uint8  & Internal tracer index; set to \texttt{255} for all rows in this release. Use \texttt{TRACERTYPE} instead. \\
\texttt{TRACERTYPE} & string & Tracer (\texttt{BGS\_ANY}, \texttt{LRG}, \texttt{ELG}, \texttt{QSO}). \\
\cline{1-3}
\end{tabular}
\end{table}


\subsection{\texttt{probabilities.tar.gz}: web-type probabilities for real objects}
\label{app:prob}

This folder contains 20 compressed FITS files (\texttt{.fits.gz}), one per rosette zone, following the same naming convention as the previous products:
\texttt{zone\_00\_probability.fits.gz}, \texttt{zone\_01\_probability.fits.gz}, \ldots, \texttt{zone\_19\_probability.fits.gz}.

They provide, at the level of \emph{real} objects, the probability that a given \texttt{TARGETID} belongs to each cosmic-web type (void, sheet, filament, knot).
The probabilities are computed from the \texttt{classification} products (\autoref{app:class}) as the fraction of the 100 random realizations in which a given real \texttt{TARGETID} is assigned to each web type.

The four probabilities are non-negative and satisfy $\mathrm{PVOID}+\mathrm{PSHEET}+\mathrm{PFILAMENT}+\mathrm{PKNOT}=1$.

\autoref{tab:dict_probability} summarizes the column dictionary applicable to all \texttt{zone\_xx\_probability.fits.gz} files in this folder.

\begin{table}[h]
\centering
\caption{Column dictionary for \texttt{zone\_xx\_probability.fits.gz} in \texttt{probabilities.tar.gz}.}
\label{tab:dict_probability}
\begin{tabular}{lll}
\cline{1-3}
\textbf{Name} & \textbf{Type} & \textbf{Description} \\
\cline{1-3}
\texttt{TARGETID}   & int64   & Identifier of the real object. \\
\texttt{TRACERTYPE} & string  & Tracer (\texttt{BGS\_ANY}, \texttt{LRG}, \texttt{ELG}, \texttt{QSO}). \\
\texttt{PVOID}      & float32 & Probability of classifying as \emph{void}. \\
\texttt{PSHEET}     & float32 & Probability of classifying as \emph{sheet}. \\
\texttt{PFILAMENT}  & float32 & Probability of classifying as \emph{filament}. \\
\texttt{PKNOT}      & float32 & Probability of classifying as \emph{knot}. \\
\cline{1-3}
\end{tabular}
\end{table}

\bibliographystyle{JHEP}
\bibliography{biblio.bib}

\end{document}